\documentclass[12pt]{article}
\usepackage{epsf}
\title{ {\bf The $\mu\rightarrow e\gamma$ and
$\tau\rightarrow \mu\gamma$ decays with the inclusion of Lorentz
and CPT violating effects}}

\author{\vspace{1cm}\\
        {\bf E. O. Iltan}
        \thanks{E-mail address:
        eiltan@heraklit.physics.metu.edu.tr}
 \\
        Physics Department, Middle East Technical University \\
        Ankara, Turkey\\}
\date{}
\begin{document}
\setlength{\baselineskip}{24pt}
\maketitle
\setlength{\baselineskip}{7mm}
\begin{abstract}
We study the Lorentz and CPT violating effects on the branching
ratio and the CP violating asymmetry of the  lepton flavor
violating interactions $\mu\rightarrow e\gamma$ and
$\tau\rightarrow \mu\gamma$, in the model III version of the two
Higgs doublet model. Here we consider that the Lorentz and CPT
violating effects exist in the QED part of the interactions and
enter into expressions  in the lepton propagators and in the
lepton-photon vertex. We observe that there exists a non-zero CP
asymmetry. However, the Lorentz and CPT violating effects on the
braching ratio and the CP asymmetry are negligibly small.
\end{abstract}
\thispagestyle{empty}
\newpage
\setcounter{page}{1}
\section{Introduction}
The Lorentz and CPT symmetries are conserved in the standard model
(SM). At higher scales, like the Planck scale, there are signals
that these symmetries are broken \cite{Kos1}, for instance in the
string theory \cite{Kos2}, in the non-commutative theories
\cite{Carroll}. In \cite{Alan} it was emphasized that the
spacetime-varying coupling constants can be associated with
violations of local Lorentz invariance and CPT symmetry. The
discussions on CPT violation on neutrino oscillations are
presented in \cite{Barenboim}.

At the low energy level, the small violations of these symmetries
can appear and with the inclusion of them the general Lorentz and
CPT violating extension of the SM is obtained \cite{Colladay,
Lehnert}. In the extension of the SM the Lorentz and CPT violating
effects are carried by the coefficients coming from an underlying
theory at the Planck scale \cite{Kos2,Carroll}. There are various
studies on the bounds of these coefficients in the literature.
They have been constrained by the experiments involving hadrons
\cite{KTeV,Opal,Kos3}, protons and neutrons \cite{Hunt}, electrons
\cite{Dehmelt, Russell}, photons \cite{Carroll2}, muons
\cite{Hughes}. The natural suppression scale for these
coefficients can be taken as the ratio of the light one $m_l$ to
the one of the order of the Planck mass . Therefore the
coefficients which carry the Lorentz and CPT violating effects are
in the the range of $10^{-23}-10^{-17}$ \cite{Russell}. Here the
first (second) number represent the electron mass $m_e$
($m_{EW}\sim 250\,GeV$) scale.

In the recent work \cite{Kos4} the amplitude for vacuum photon
splitting in the framework of general Lorentz and CPT violating
Quantum Electro Dynamics (QED) extensions have been analyzed and
it was observed that radiative corrections arising from Lorentz
violation in the fermion sector induce the vacuum photon
splitting. In \cite{Kos5} the one loop renormalizability of the
general Lorentz and CPT violating extension QED has been showed.

In the present work we study the Lorentz and CPT violating effects
on the branching ratio $(BR)$ and the CP violating asymmetry
$A_{CP}$ for the  lepton flavor violating (LFV) interactions
$\mu\rightarrow e\gamma$ and $\tau\rightarrow \mu\gamma$, in the
model III version of the two Higgs doublet model (2HDM), since
these decays do not exist in the SM. Here we consider that the
Lorentz and CPT violating effects exist in the QED part of the
interactions and enter into expressions  in the lepton propagators
and the lepton-photon vertex.

In the literature, there are several studies on LFV interactions
in different models. Such interactions are analyzed in a model
independent way in \cite{Chang}, in the framework of model III
2HDM \cite{Eilt1,Eilt2, Diaz}, in supersymmetric models
\cite{Barbieri1,Barbieri2,Barbieri3,Ciafaloni,Duong,Couture,Okada}.
Furthermore the experimental current limits for the  BR's ratios
of the processes $\mu\rightarrow e\gamma$ and $\tau\rightarrow
\mu\gamma$ are $1.2\times 10^{-11}$ \cite{Brooks} and $1.1\times
10^{-6}$ \cite{Ahmed} respectively.

The inclusion of the Lorentz and CPT violating effects in the
model III  does not bring a detectable correction to the BR of the
LFV processes under consideration, since the corresponding
coefficients are highly suppressed at the low energy scale.
However we try to examine the relative  importance of the
different coefficients which switch on the Lorentz and CPT
violating effects in the BR of the decays $\mu\rightarrow e\gamma$
and $\tau\rightarrow \mu\gamma$. In addition to this, we analyze
the possible $A_{CP}$ in these decays and the coefficients which
are sources of $A_{CP}$. Notice that  the $A_{CP}$ does not exist
for the LFV decays $l_1\rightarrow l_2\gamma$ in the framework of
the model III and the possibility of such asymmetry has been
studied in \cite{Eilt2}. In this work it was assumed that the
$A_{CP}$ could be switched on when one considered  the model
beyond the model III and insert a new parameter into the
interactions. The magnitude of the $A_{CP}$ is directly
proportional to this new parameter. The inclusion of the Lorentz
and CPT violating effects in the model III causes a non-zero
$A_{CP}$, however it is too small to be detected, similar to the
corrections on the BRs of these processes.

The paper is organized as follows: In Section 2, we present the
theoretical expression for the matrix element and the $A_{CP}$ of
LFV interaction $l_1\rightarrow l_2\gamma$ with the inclusion of
the Lorentz and CPT violating effects. Section 3 is devoted to
discussion and our conclusions. In the Appendix, we present the
explicit forms of the functions appearing in the calculation of
the matrix element of the decays under consideration.
\section{The LFV interactions $\mu\rightarrow e\gamma$ and
$\tau\rightarrow \mu\gamma$ with the addition of Lorentz and CPT
violating effects }
This section is devoted to the derivation of the Lorentz and CPT
violating effects on the $BR$ and the CP asymmetry  of the LFV
$l_1\rightarrow l_2\gamma$ decay. The LFV interactions in the tree
level are allowed in the general 2HDM, the so-called model III and
the the LFV process can be regulated by the Yukawa interaction,
\begin{eqnarray}
{\cal{L}}_{Y}= \eta^{E}_{ij} \bar{l}_{i L} \phi_{1} E_{j R}+
\xi^{E}_{ij} \bar{l}_{i L} \phi_{2} E_{j R} + h.c. \,\,\, ,
\label{lagrangian}
\end{eqnarray}
where $i,j$ are family indices of leptons, $L$ and $R$ denote
chiral projections $L(R)=1/2(1\mp \gamma_5)$, $\phi_{i}$ for
$i=1,2$, are the two scalar doublets, $l_{i L}$ and $E_{j R}$ are
lepton doublets and singlets respectively.
Here $\phi_{1}$ and $\phi_{2}$ are chosen as
\begin{eqnarray}
\phi_{1}=\frac{1}{\sqrt{2}}\left[\left(\begin{array}{c c}
0\\v+H^{0}\end{array}\right)\; + \left(\begin{array}{c c}
\sqrt{2} \chi^{+}\\ i \chi^{0}\end{array}\right) \right]\, ;
\phi_{2}=\frac{1}{\sqrt{2}}\left(\begin{array}{c c}
\sqrt{2} H^{+}\\ H_1+i H_2 \end{array}\right) \,\, ,
\label{choice}
\end{eqnarray}
where only $\phi_{1}$ has a vacuum expectation value;
\begin{eqnarray}
<\phi_{1}>=\frac{1}{\sqrt{2}}\left(\begin{array}{c c}
0\\v\end{array}\right) \,  \, ;
<\phi_{2}>=0 \,\, .
\label{choice2}
\end{eqnarray}
Now we consider the gauge and $CP$ invariant Higgs potential which
spontaneously breaks  $SU(2)\times U(1)$ down to $U(1)$  as:
\begin{eqnarray}
V(\phi_1, \phi_2 )&=&c_1 (\phi_1^+ \phi_1-v^2/2)^2+ c_2 (\phi_2^+
\phi_2)^2 \nonumber \\ &+&  c_3 [(\phi_1^+ \phi_1-v^2/2)+ \phi_2^+
\phi_2]^2 + c_4 [(\phi_1^+ \phi_1)
(\phi_2^+ \phi_2)-(\phi_1^+ \phi_2)(\phi_2^+ \phi_1)] \nonumber \\
&+& c_5 [Re(\phi_1^+ \phi_2)]^2 + c_{6} [Im(\phi_1^+ \phi_2)]^2
+c_{7} \, , \label{potential}
\end{eqnarray}
with constants $c_i, \, i=1,...,7$. With the choice of $\phi_{1}$,
$\phi_{2}$ and the potential $V(\phi_1, \phi_2)$,  $H_1$ and $H_2$
are obtained as the mass eigenstates $h^0$ and $A^0$ respectively,
since no mixing occurs between two CP-even neutral bosons $H^0$
and $h^0$ in the tree level.

The FCNC is produced by the part of the lagrangian
\begin{eqnarray}
{\cal{L}}_{Y,FC}= \xi^{E}_{ij} \bar{l}_{i L} \phi_{2} E_{j R} +
h.c. \,\, . \label{lagrangianFC}
\end{eqnarray}
Here the Yukawa matrices $\xi^{E}_{ij}$ are responsible for the
LFV interactions and, in general, they have complex entries.
Notice that in the following we replace $\xi^{E}$ with
$\xi^{E}_{N}$ where "N" denotes the word "neutral".

At this stage we insert the Lorentz and CPT violating effects with
the assumption that they exist in the QED part of the
interactions. The fermionic part of the general Lorentz and CPT
violating QED lagrangian in 4 space-time dimensions reads
\cite{Colladay}
\begin{eqnarray}
L=\frac{i}{2} \bar{\psi} \Gamma^{\mu} D_{\mu} \psi-\bar{\psi} M
\psi \label{CPTLag}
\end{eqnarray}
where
\begin{eqnarray}
\Gamma^{\mu}&=&\gamma^{\mu}+\Gamma_1^{\mu} \nonumber \\
M&=&m+m_1 \label{GammaM1}
\end{eqnarray}
and
\begin{eqnarray}
\Gamma_1^{\mu}&=&c^{\alpha\mu}\,\gamma_{\alpha}+d^{\alpha\mu}\,\gamma_5\,
\gamma_{\alpha}+e^{\mu}+i
f^{\mu}\,\gamma_5+\frac{1}{2}\,g^{\lambda\nu\mu}\,
\sigma_{\lambda\nu}  \, ,\nonumber \\
m_1&=& a_{\mu}\, \gamma^{\mu}+b_{\mu} \gamma_{5}\,
\gamma^{\mu}+\frac{1}{2}\, h_{\mu\nu} \sigma^{\mu\nu}.
\label{GammaM2}
\end{eqnarray}
Here the coefficients $a_{\mu}$, $b_{\mu}$, $c_{\alpha \mu}$,
$d_{\alpha\mu}$, $e_{\mu}$, $f_{\mu}$, $g_{\lambda\nu\mu}$ and
$h_{\mu\nu}$ cause the Lorentz violation. Among them $a_{\mu}$,
$b_{\mu}$, $e_{\mu}$, $f_{\mu}$ and $g_{\lambda\nu\mu}$ are
responsible for the CPT violation (see \cite{Kos5} for details).

\begin{figure}[htb]
\vskip 0.0truein \centering \epsfxsize=6.8in
\leavevmode\epsffile{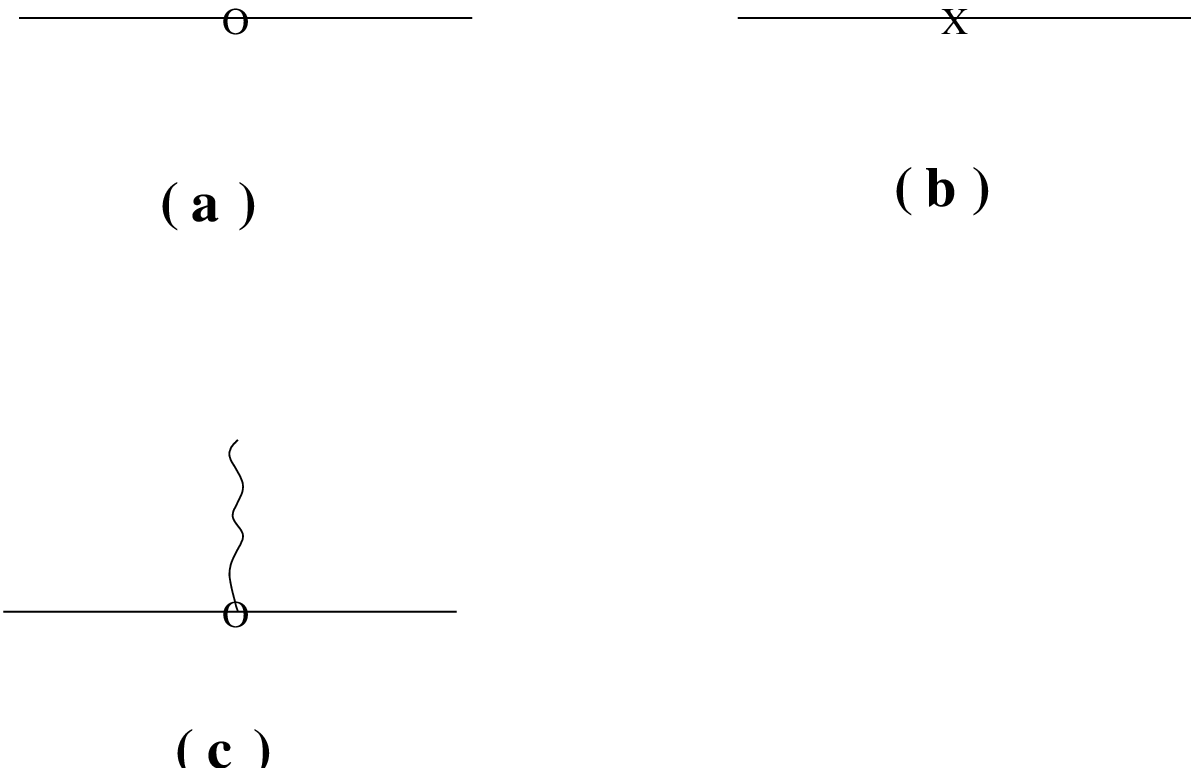} \vskip 1.0truein \caption[]{The CPT
and Lorentz violating insertions. {\bf (a) }. lepton propagator
insertion ($= i\,\Gamma_1^{\mu} p_{\mu}$ where $p$ is the
4-momentum vector of the internal lepton). {\bf (b)}. lepton
propagator mass insertion ($= -i\,m_1$) {\bf (c)}. lepton-photon
insertion ($= -i q\,\Gamma_1^{\mu}$ where $q$ is the lepton
charge).} \label{fig1}
\end{figure}

\begin{figure}[htb]
\vskip -0.4 truein \centering \epsfxsize=6.8in
\leavevmode\epsffile{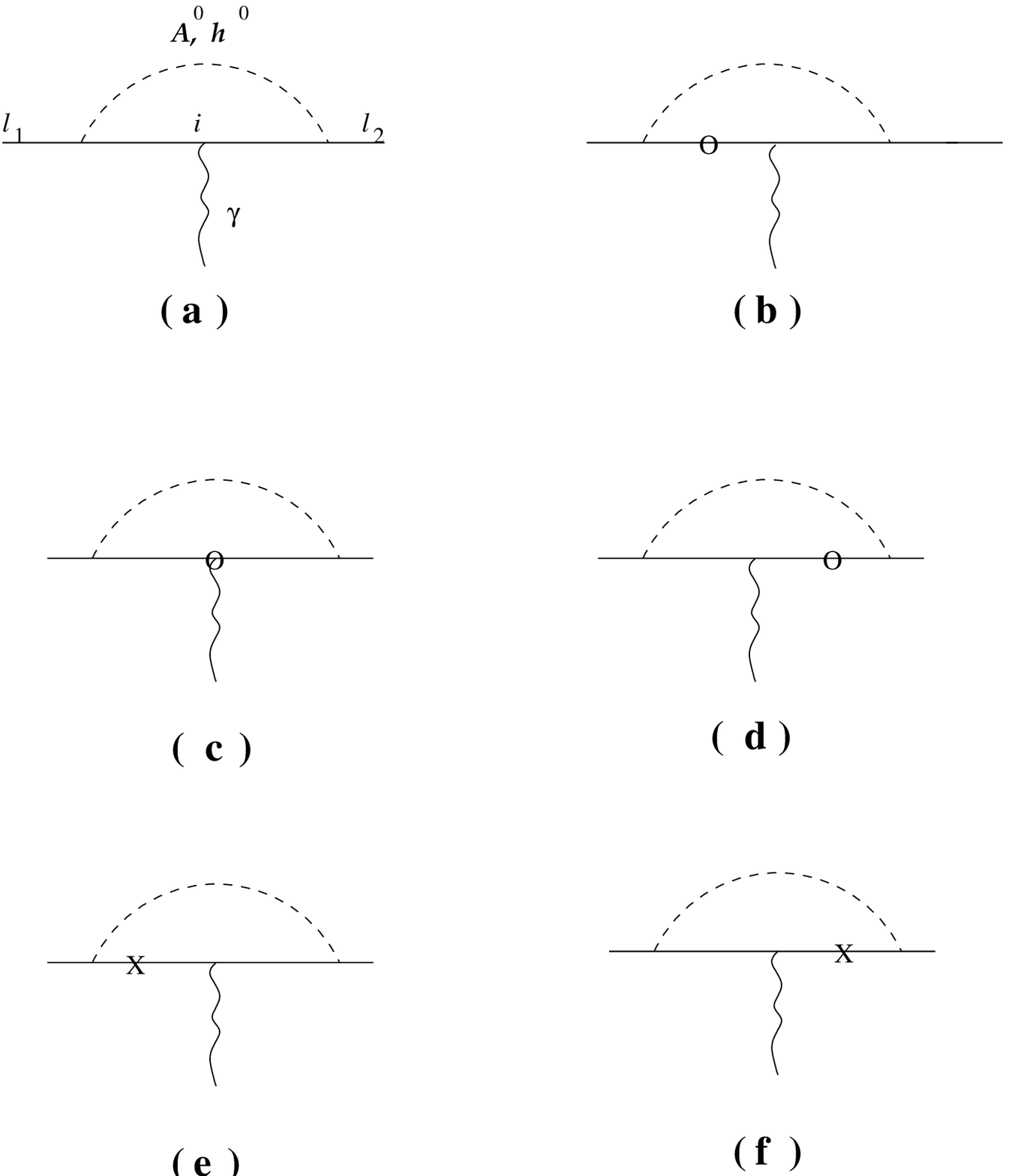} \vskip 1.0truein \caption[]{One loop
diagrams contribute to the LFV interactions $l_1\rightarrow l_2
\gamma$ due to the neutral Higgs bosons $h_0$ and $A_0$ in the
2HDM. Dashed (curly, straight) lines represent $h_0$ and $A_0$
fields (electromagnetic field, lepton), the signs $O$ and $\times$
represent the insertions into the propagator and the
photon-fermion vertex.} \label{fig2}
\end{figure}

The LFV $l_1\rightarrow l_2 \gamma$ interaction occurs with the
help of the neutral Higgs bosons, namely, Higgs bosons $h_0$ and
$A_0$, in the model III. The Lorentz and CPT  violating effects in
$\Gamma_1^{\mu}$ and $m_1$ (see eq. \ref{GammaM2}) are taken into
account with the insertions in the internal lepton propagator and
the additional fermion-photon vertex (see Fig. \ref{fig1}). Now we
use the on-shell renormalization scheme to calculate the matrix
element for the LFV process under consideration. Since, in the
on-shell renormalization scheme, the self energy $\sum(p)$ can be
written as
\begin{eqnarray}
\sum(p)=(\hat{p}-m_{l_2})\bar{\sum}(p) (\hat{p}-m_{l_1})\,\, ,
\label{self}
\end{eqnarray}
with $\hat{p}=\gamma_{\mu} p^{\mu}$, the corresponding diagrams
vanish when $l_1 (l_2)$-lepton is on-shell. However, the vertex
diagrams (Fig. \ref{fig2}) give a non-zero contribution and the
logarithmic divergences are eliminated by inserting an appropriate
the counter term $\Gamma^C$
\begin{eqnarray}
\Gamma^{Ren}_{\mu}=\Gamma^0_{\mu}+\Gamma^C_{\mu}  \ ,
,\label{GammaRen2}
\end{eqnarray}
where $\Gamma^{Ren}_{\mu}$ and $\Gamma^0_{\mu}$ are renormalized
and bare vertex functions respectively. Here the
$\Gamma^{Ren}_{\mu}$ satisfies the equation,
\begin{eqnarray}
k^{\mu}\,\Gamma^{Ren}_{\mu}=0 \, . \label{GammaRen}
\end{eqnarray}
where $k$ is the 4-momentum vector of outgoing photon. Now, the
matrix element of the LFV $l_1\rightarrow l_2 \gamma$ process with
the addition of the Lorentz and CPT violating effects is obtained
as,
\begin{eqnarray}
M&=& \frac{\sqrt{4 \pi \alpha_e}}{32 \pi^2} \,\Bigg \{ A
\gamma_{\mu}+B_{\mu \alpha} \gamma_{\alpha}+i H_{\alpha}
\sigma_{\mu\alpha}+E_{\mu} \nonumber
\\ &+& A^{\prime} \gamma_{\mu}\gamma_{5}+ B^{\prime}_{\mu \alpha}
\gamma_{\alpha} \gamma_{5}+H^{\prime}_{\alpha}
\sigma_{\mu\alpha}\gamma_{5}+E^{\prime}_{\mu}\gamma_{5} \Bigg \}
\, . \label{Mat}
\end{eqnarray}
where $\alpha_e=\frac{1}{137}$ and
\begin{eqnarray}
A&=& \sum_{i=e,\mu,\tau}\, Q_i \, \xi_{N,i l_1} \, \xi_{N,i l_2}
\int_0^1\, dx \int_0^{1-x}\, dy \, (1-x-y)\, A_i(x,y) \, , \nonumber \\
B_{\mu \alpha}&=& \sum_{i=e,\mu,\tau} Q_i \, \xi_{N,i l_1}
\xi_{N,i l_2}\int_0^1\, dx\int_0^{1-x}\,dy \, (1-x-y)\, B_{i\,\mu
\alpha}(x,y) \, ,\nonumber \\
H_{\alpha}&=& \sum_{i=e,\mu,\tau} Q_i \,\xi_{N,i l_1} \xi_{N,i
l_2} \int_0^1\, dx\int_0^{1-x}\,dy \, (1-x-y)\,
H_{i\,,\alpha}(x,y) \, ,
\nonumber \\
E_{\mu}&=& \sum_{i=e,\mu,\tau} Q_i \,\xi_{N,i l_1} \xi_{N,i l_2}
\int_0^1\, dx\int_0^{1-x}\,dy \, (1-x-y)\, E_{i\,,\mu}(x,y) \, , \nonumber \\
A^{\prime}&=& \sum_{i=e,\mu,\tau} Q_i \,\xi_{N,i l_1} \xi_{N,i
l_2} \int_0^1\, dx\int_0^{1-x}\,dy \, (1-x-y)\, A^{\prime}_i(x,y) \, ,
\nonumber \\
B^{\prime}_{\mu \alpha}&=& \sum_{i=e,\mu,\tau} Q_i \xi_{N,i l_1}
\xi_{N,i l_2} \int_0^1\, dx\int_0^{1-x}\,dy \, (1-x-y)\,
B^{\prime}_{i\,\mu \alpha}(x,y) \, , \nonumber
\\H^{\prime}_{\alpha}&=& \sum_{i=e,\mu,\tau} Q_i \,\xi_{N,i l_1}
\xi_{N,i l_2} \int_0^1\, dx\int_0^{1-x}\,dy \, (1-x-y)\,
H^{\prime}_{i\,,\alpha}(x,y) \, , \nonumber \\
E^{\prime}_{\mu}&=& \sum_{i=e,\mu,\tau} Q_i \,\xi_{N,i l_1}
\xi_{N,i l_2}\int_0^1\, dx\int_0^{1-x}\,dy \, (1-x-y)\,
E^{\prime}_{i\, ,\mu}(x,y) \, , \label{Mat2}
\end{eqnarray}
and
\begin{eqnarray}
H_{i\, ,\alpha}(x,y)&=&C_{i\,,\alpha}(x,y)+ D_{i}(x,y)\,
k_{\alpha}
 \, , \nonumber \\
H^{\prime}_{i\,,\alpha}(x,y)&=&C^{\prime}_{i\,,\alpha}(x,y)+
D^{\prime}_{i}(x,y)\, k_{\alpha} \, , \label{Hh}
\end{eqnarray}
Here  $l_1$ ($l_2$) is incoming (outgoing) leptons, i, $Q_i$
denote the internal leptons ($i=e,\mu,\tau$), their charges and we
choose the Yukawa coupling $\xi_{N,i l_{1 (2)}}$ real. In eqns.
(\ref{Mat2}) and (\ref{Hh}) the coefficients
$A^{(\prime)}_i(x,y)$, $B^{(\prime)}_{i\,\mu \alpha}(x,y)$,
$C^{(\prime)}_{i\,,\alpha}(x,y)$, $D^{(\prime)}_{i}(x,y)$,
$E^{(\prime)}_{i\,,\mu}(x,y)$ are given in the Appendix.

Finally, using the well known expression,
\begin{equation}
d\Gamma=\frac{(2\, \pi)^4}{2\,m_{l_1}} \, |M|^2\,\delta^4
(p-\sum_{i=1}^2 p_i)\,\prod_{i=1}^2\,\frac{d^3 p_i}{(2 \pi)^3\, 2
E_i} \,
 ,\label{DecWidth}
\end{equation}
the decay width $\Gamma$ is obtained in the $l_1$ lepton rest
frame. Here $p$ ($p_i$, i=1,2) is four momentum vector of $l_1$
lepton, ($l_2$ lepton, outgoing $k$ photon).

At this stage, we calculate the CP asymmetry $A_{CP}$ of the
process $l_1 \rightarrow l_2 \gamma$. In the model III the CP
violation does not exist even with the choice of complex Yukawa
couplings $\xi_{N,ij}$. However the addition of the Lorentz and
CPT violating terms switch on the CP violating effects and these
effects are very small since they are proportional to the the
Lorentz and CPT violating coefficients. Using the definition of
the CP asymmetry $A_{CP}$
\begin{eqnarray}
A_{CP}=\frac{\Gamma - \bar{\Gamma}}{\Gamma + \bar{\Gamma}} \, ,
\label{ACP1}
\end{eqnarray}
where $\bar{\Gamma}$ denotes the decay width for the CP conjugate
process, we get
\begin{eqnarray}
A_{CP}=\frac{\int_{0}^{1} dx \int_{0}^{1-x} dy \,\Omega
(x,y)}{\int_{0}^{1} dx \int_{0}^{1-x} dy \,W(x,y)} \label{ACP}
\end{eqnarray}
where
\begin{eqnarray}
\Omega
(x,y)=\Omega_g(x,y)+\Omega_a(x,y)+\Omega_b(x,y)+\Omega_e(x,y)
\label{omega1}
\end{eqnarray}
with
\begin{eqnarray}
\Omega_g(x,y)&=&
-\frac{4\pi\alpha_e}{512\,\pi^4}\,\sum_{i=e,\mu,\tau}
m_i\,m_{l_1}^4\, Q_i^2\, |\xi_{N,il_2}|^2\, |\xi_{N,il_1}|^2\,
sin\,2\,\theta_{i l_1}\, g[p_0,k,k]\, (F(z_A)-F(z_h))\nonumber
\\ &\times& (1-x-y)\, G_1(x,y)\, ,
\nonumber \\
\Omega_a (x,y)&=&
-\frac{4\pi\alpha_e}{384\,\pi^4}\,\sum_{i=e,\mu,\tau}
\,m_{l_1}^4\, Q_i^2\, |\xi_{N,il_2}|^2\, |\xi_{N,il_1}|^2\,
cos^2\,\theta_{i l_1}\, a[p_0]\, (H_1(z_h)-H_2(z_A))\nonumber
\\ &\times& (1-x-y)\, G_2(x,y)\, ,
\nonumber \\
\Omega_b (x,y)&=&
-\frac{4\pi\alpha_e}{384\,\pi^4}\,\sum_{i=e,\mu,\tau}
\,m_{l_1}^4\, m_i\, Q_i^2\, |\xi_{N,il_2}|^2\, |\xi_{N,il_1}|^2\,
cos^2\,\theta_{i l_1}\, b[p_0]\, (H_1(z_h)-H_2(z_A))\nonumber
\\ &\times& (1-x-y)\, G_3(x,y)\, ,
\nonumber \\
\Omega_e (x,y)&=&
-\frac{4\pi\alpha_e}{384\,\pi^4}\,\sum_{i=e,\mu,\tau}
\,m_{l_1}^4\, Q_i^2\, |\xi_{N,il_2}|^2\, |\xi_{N,il_1}|^2\,
cos^2\,\theta_{i l_1}\, e[p_0]\, (H_1(z_h)-H_2(z_A))\nonumber
\\ &\times& (1-x-y)\, G_4(x,y)\, , \label{omega2}
\end{eqnarray}
Here the function $F(z_S)$, $H_i (x,y)$ and $G_i (x,y)$ read
\begin{eqnarray}
F(z_S)&=& \frac{1}{m_S^2\,(z_S-1)^3}\,
(3-4\,z_S+z_S^2+2\,ln\,z_S) \, , \nonumber \\
H_1(z_S)&=& \frac{1}{m_S^2\,(z_S-1)^4}\, \Bigg(
(z_S-1)\,\Big(6\,m_i\,(3-4\,z_S+z_S^2)+m_{l_1}\, (-4+5\,z_S+5\,
z_S^2)\Big) \nonumber \\ &+& 6 \,\Big(
2\,m_i\,(z_S-1)+m_{l_1}\,(1-2\,z_S)\,z_S
\Big)\,ln\,z_S \Bigg)\, , \nonumber \\
H_2(z_S)&=& -6\,m_i\,F(z_S)
\, , \nonumber \\
G_1(x,y)&=& (\frac{1}{L_{1,h^0}}-\frac{1}{L_{1,A^0}})
\,x\,y-(\frac{1}{L_{2,h^0}}-\frac{1}{L_{2,A^0}}) \,x\,(x+y-2)
\nonumber \\ &+& 2\,(\frac{1}{L^2_{2,h^0}}+\frac{1}{L^2_{2,A^0}})
m_i\,m_{l_1}\,x^2\,(x+y)
\, , \nonumber \\
G_2(x,y)&=& \frac{m_i+m_{l_1}\,x}{L^2_{2,h^0}}
-\frac{m_i-m_{l_1}\,x}{L^2_{2,A^0}}
\, , \nonumber \\
G_3(x,y)&=& \frac{x\,(1-4\,x-4\,y)}{4}\, (\frac{1}{L^2_{1,h^0}}
-\frac{1}{L^2_{1,A^0}})\,
\, , \nonumber \\
G_4(x,y)&=& \frac{x\, (3-4\,x-4\,y)}{2}\,(\frac{1}{L_{2,h^0}}
-\frac{1}{L_{2,A^0}}+\frac{1}{L_{1,h^0}}
-\frac{1}{L_{1,A^0}})-x\,\Bigg(
\frac{f_1^+(x,y)}{L^2_{2,A^0}}+\frac{f_1^-(x,y)}{L^2_{2,h^0}}
\nonumber \\
&+& \frac{f_2^+(x,y)}{L^2_{1,A^0}}+\frac{f_2^-(x,y)}{L^2_{1,h^0}}
\Bigg) \, , \label{Gxy}
\end{eqnarray}
and
\begin{eqnarray}
f_1^+ (x,y)&=&2
\,m_i\,m_{l_1}\,x+m_i^2\,(-1+x-y)-m^2_{l_1}\,x\,y\,(-1+x+y)
\, , \nonumber \\
f_1^- (x,y)&=&2
\,m_i\,m_{l_1}\,x+m_i^2\,(1-x+y)+m^2_{l_1}\,x\,y\,(-1+x+y)
\, , \nonumber \\
f_2^+ (x,y)&=&2
\,m_i\,m_{l_1}\,x+m_i^2\,(-1+x-y)+m^2_{l_1}\,x\,(x+y)\,(-1+x+y)
\, , \nonumber \\
f_2^- (x,y)&=&2
\,m_i\,m_{l_1}\,x+m_i^2\,(1-x+y)+m^2_{l_1}\,x\,(x+y)\,(1-x-y) \, .
\label{fxy}
\end{eqnarray}
Notice that the index $k=1,2,3$; the parameters $L_{(1,2),S}$,
$z_S$ are given in the Appendix and we do not give the explicit
expression for the function $W(x,y)$ since it is very long. Here
we take the Yukawa couplings $\xi_{N,il_1}$ complex with the
parametrization
\begin{eqnarray}
\xi^{E}_{N,il_1}=|\xi^{E}_{N,i l_1}|\, e^{i\theta_{il_1}} \, ,
\label{ksipar}
\end{eqnarray}
and  $\xi_{N,il_2}$ real. In eq. (\ref{omega2}), the coefficients
$g[p_0,k,k]=m_{l_1}\, g[0,k,k]$, $a[p_0]=m_{l_1}\,a[0]$,
$b[p_0]=m_{l_1}\,b[0]$ and $e[p_0]=m_{l_1}\,e[0]$ exist since we
study in the rest frame of the incoming lepton $l_1$. Here
$g[0,k,k]$ is CP even and the others, $a[0],b[0],e[0]$, are CP odd
coefficients appearing in the lagrangian eq. (\ref{CPTLag}) (see
\cite{Kos1}). The $A_{CP}$ is nonzero due to the complex nature of
the couplings for the part $\Omega_g$ and CP odd nature of the
coefficients for the part $\Omega_a+\Omega_b+\Omega_e$.
\section{Discussion}
In this section we analyze the Lorentz and CPT violating  effects
on the $BR$ and  the  $A_{CP}$ for the LFV $\mu\rightarrow
e\gamma$ and $\tau\rightarrow \mu\gamma$ decays in the framework
of the model III. The Yukawa couplings $\xi^{E}_{N,i\mu}$ and
$\xi^{E}_{N,i\tau}$ $(i=e,\mu,\tau)$ are responsible for the LFV
decays. They are the free parameters of the theory and they should
be restricted by respecting the appropriate experimental
measurements. Fortunately, the strength of the Yukawa couplings
$\xi^{E}_{N,i j}$ are considered as proportional to the masses of
the leptons which are given by the indices and therefore, the
contribution of the couplings related to the heavy leptons are
dominant. For the decays $\mu\rightarrow e\gamma$ and $\tau
\rightarrow \mu\gamma$ the main contribution comes from the
internal $\tau$ lepton and the Yukawa couplings
$\xi^{E}_{N,\tau\mu}$, $\xi^{E}_{N,\tau e}$  and
$\xi^{E}_{N,\tau\tau}$, $\xi^{E}_{N,\tau\mu}$ play the main role
respectively. There are various studies on the strength of these
couplings in the literature. The upper limit of the coupling
$\xi^{E}_{N,\tau\mu}$ has been predicted as $\sim 0.15$, by using
experimental result of anomalous magnetic moment of muon in
\cite{Erilano}. In \cite{IltSund} the coupling
$\xi^{E}_{N,\tau\mu}$ and $\xi^{E}_{N,\tau\tau}$ has been
estimated at the order of $0.03$ and $0.15$ respectively. For the
coupling $\xi^{E}_{N,\tau e}$ the prediction has been done at the
order of the magnitude of $10^{-4}-10^{-3}$ in \cite{Eilt1}, by
using the experimental result of the electric dipole moment of
muon \cite{Abdullah} and the upper limit of the $BR$ of the
process $\mu\rightarrow e\gamma$ \cite{Mega}. Notice that the
couplings $\xi^{E}_{N,ij}$ are complex in general and in the
following, we use the parametrization
\begin{equation}
\xi^{E}_{N,ij}= \sqrt{\frac{4 G_F}{\sqrt {2}}}
\bar{\xi}^{E}_{N,ij} \, , \label{ksipar2}
\end{equation}
where $G_F=1.6637 \times 10^{-5} (GeV^{-2})$ is the fermi
constant.

A possible small violations of Lorentz and CPT symmetry in the
extension of the SM  arise and those effects could be detected in
the existing experiments \cite{Kos1}. In the present work we
assume that these effects are only due to the QED part of the
interactions and we take their effects into account. Even if their
contributions are negligible in our processes, we try to
understand the relative behavior of different coefficients,
violating Lorentz and CPT symmetry, in the $BR$ and the CP
violating asymmetry $A_{CP}$. In our calculations we take the
numerical values of the coefficients at the order of the magnitude
of $10^{-20}-10^{-18}$ respecting the existing results \cite{Kos5}
\begin{eqnarray}
|a|, |b| &\sim& m_{\mu} (m_{\tau}) \frac{m_{EW}}{M_P} <
10^{-18}\, (10^{-17}) \, GeV \, , \nonumber \\
|d|, |c|,|e|, |f|,|g|&\sim&  \frac{m_{\mu}\, (m_{\tau})}{M_P}
< 10^{-21}\, (10^{-20}) \, , \nonumber \\
\label{coeff}
\end{eqnarray}
where $m_{EW}$ (${M_P}$) is the electro weak (Planck mass) scale.

\begin{figure}[htb]
\vskip -3.0truein \centering \epsfxsize=6.8in
\leavevmode\epsffile{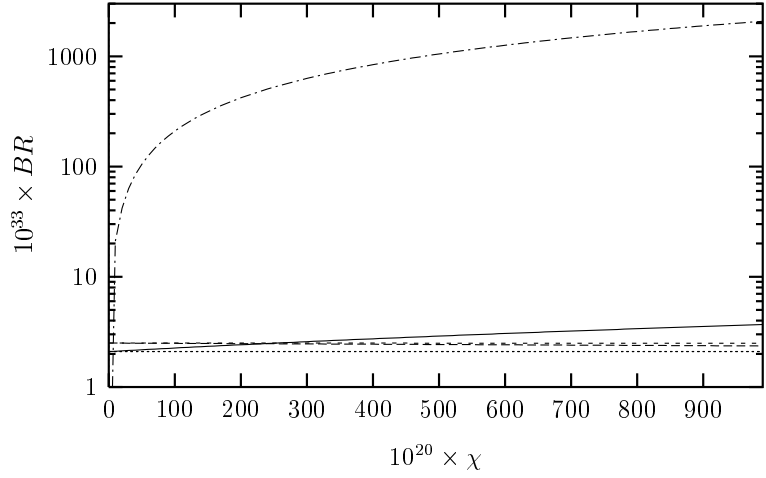} \vskip -3.0truein \caption[]{
The magnitude $\chi$ of the coefficient dependence of the Lorentz
violating part of the $BR$ for the decay $\mu\rightarrow e\gamma$,
for the real Yukawa couplings, $\bar{\xi}^{E}_{N,\tau\mu}=30\,
GeV$, $\bar{\xi}^{E}_{N,\tau e}=0.001\, GeV$. Here solid (dashed,
small dashed,dotted, dot-dashed) line represents the dependence to
the coefficient $|a.p|$ ($|b.p|$, ($c^{Sym}$, $d^{Sym}$,
$|g[p,\beta,\beta]|$), $|e.p|$, $|f.p|$), in the case that the
other coefficients have the same numerical value $10^{-20}$. }
\label{Brmuegamma}
\end{figure}

In Fig. \ref{Brmuegamma}, we present the magnitude of the
coefficient dependence of the Lorentz violating part of the $BR$
for the decay $\mu\rightarrow e\gamma$, for the real Yukawa
couplings, $\bar{\xi}^{E}_{N,\tau\mu}=30\, GeV$,
$\bar{\xi}^{E}_{N,\tau e}=0.001\, GeV$ . Here solid (dashed, small
dashed,dotted, dot-dashed) line represents the dependence to the
coefficient $|a.p|$, $|b.p|$, $c^{Sym}$, ($d^{Sym}$, $|f.p|$,
$|g[p,\beta,\beta]|$), $|e.p|$), in the case that the other
coefficients have the same numerical value $10^{-20}$. The BR is
at the order of the magnitude of $10^{-33}-10^{-30}$, which is a
negligible quantity compared to the current experimental limits
$10^{-11}$ \cite{Brooks}. It increases with the increasing values
of the coefficients, especially $|f|$ and $|a|$. The BR is much
more sensitive to the coefficient $|f|$ compared to the others.

\begin{figure}[htb]
\vskip -3.0truein \centering \epsfxsize=6.8in
\leavevmode\epsffile{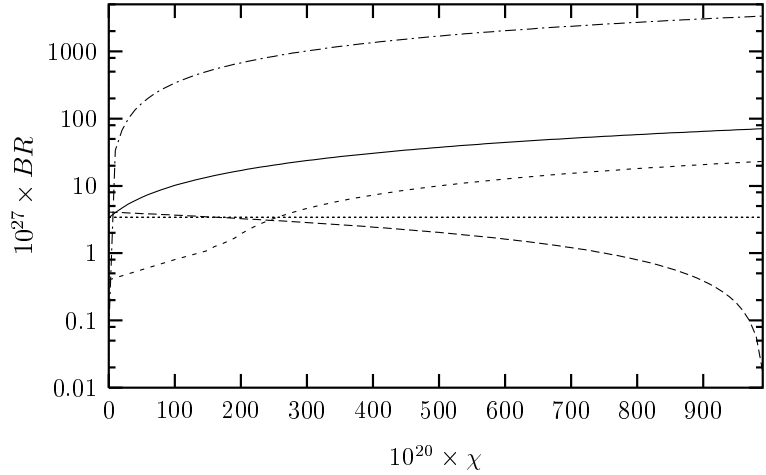} \vskip -3.0truein
\caption[]{The same as Fig. \ref{Brmuegamma} but for the decay
$\tau\rightarrow \mu\gamma$.} \label{Brtaumugamma}
\end{figure}

Fig.\ref{Brtaumugamma} is devoted to the magnitude of the
coefficient dependence of the Lorentz violating part of the $BR$
for the decay $\tau\rightarrow \mu\gamma$, for the real Yukawa
couplings, $\bar{\xi}^{E}_{N,\tau\tau}=100\, GeV$,
$\bar{\xi}^{E}_{N,\tau \mu}=30\, GeV$. Here solid (dashed, small
dashed,dotted, dot-dashed) line represents the dependence to the
coefficient $|a.p|$ ($|b.p|$, $c^{Sym}$, ($d^{Sym}$, $|f.p|$,
$|g[p,\beta,\beta]|$), $|e.p|$), in the case that the other
coefficients have the same numerical value $10^{-20}$. The BR is
at the order of the magnitude of $10^{-29}-10^{-24}$, which is a
negligible quantity compared to the current limits $10^{-6}$
\cite{Ahmed}, similar to the previous process. Here the increasing
values of the coefficients, $|f|$, $|a|$ and $c^{Sym}$ increases
the BR. However the BR decreases with the increasing values the
coefficient $|b|$. The BR is much more sensitive to the
coefficients $|e|$, $|a|$, $|e|$ and $|b|$ compared to the others.

\begin{figure}[htb]
\vskip -3.0truein \centering \epsfxsize=6.8in
\leavevmode\epsffile{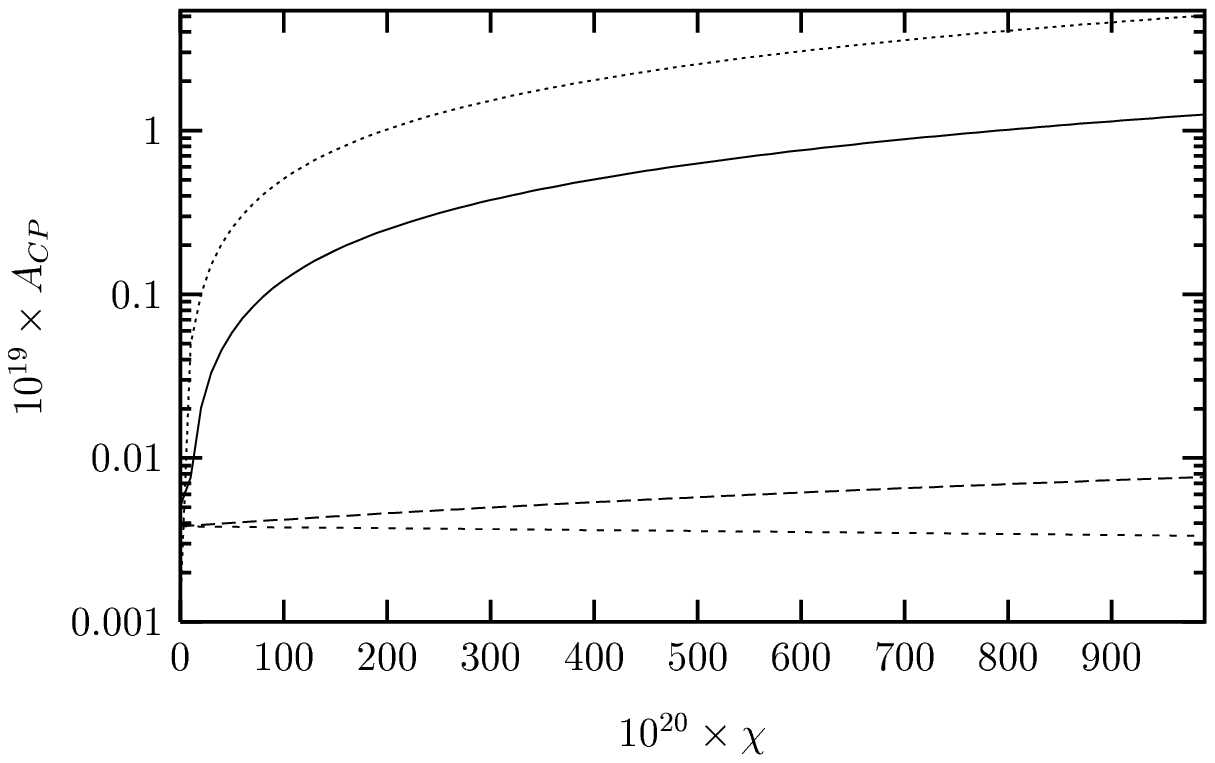} \vskip -3.0truein
\caption[]{The magnitude $\chi$ of the coefficient dependence of
the $A_{CP}$ for the decay $\mu\rightarrow e\gamma$, for the
Yukawa couplings,$|\bar{\xi}^{E}_{N,\tau\mu}|=30\, GeV$,
$|\bar{\xi}^{E}_{N,\tau e}|=0.001\, GeV$ ,
$sin\theta_{\tau\mu}=0.5$, $sin\theta_{\tau e}=0$. Here solid
(dashed, small dashed,dotted) line represents the dependence to
the coefficient $|g[p_0,k,k]|$, $|a[p_0]|$, $|b[p_0]|$ and
$|e[p_0]|$, in the case that the other coefficients have the same
numerical value $10^{-20}$.} \label{ACPmuegam}
\end{figure}

\begin{figure}[htb]
\vskip -3.0truein \centering \epsfxsize=6.8in
\leavevmode\epsffile{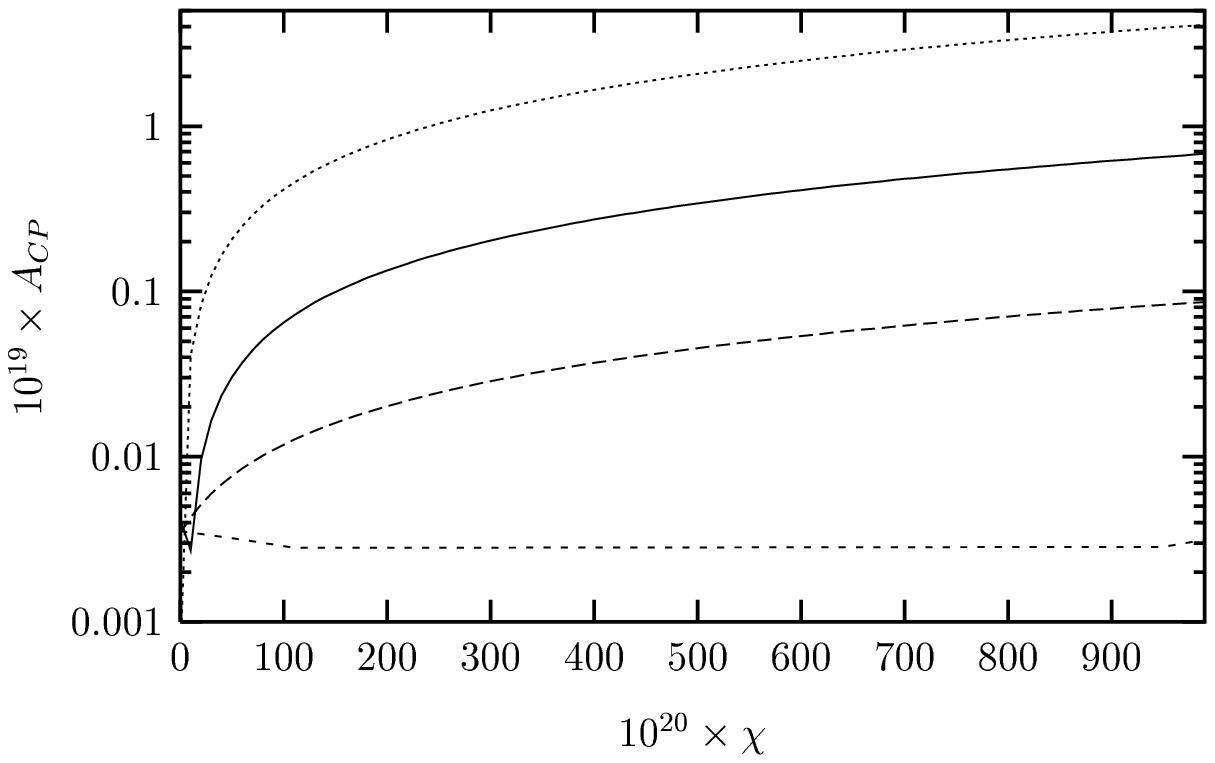} \vskip -3.0truein
\caption[]{The magnitude $\chi$ of the coefficient dependence of
the $A_{CP}$ for the decay $\tau\rightarrow \mu\gamma$, for the
Yukawa couplings,$|\bar{\xi}^{E}_{N,\tau\tau}|=100\, GeV$,
$|\bar{\xi}^{E}_{N,\tau \mu}|=30\, GeV$ ,
$sin\theta_{\tau\tau}=0.5$, $sin\theta_{\tau \mu}=0$. Here solid
(dashed, small dashed,dotted) line represents the dependence to
the coefficient $|g[p_0,k,k]|$, $|a[p_0]|$, $|b[p_0]|$ and
$|e[p_0]|$, in the case that the other coefficients have the same
numerical value $10^{-20}$.} \label{ACPtaumugam}
\end{figure}

Now, in Fig. \ref{ACPmuegam} (\ref{ACPtaumugam}), we present the
possible CP violating asymmetry $A_{CP}$ for the decay
$\mu\rightarrow e\gamma$ ($\tau\rightarrow \mu\gamma$), for the
Yukawa couplings,$|\bar{\xi}^{E}_{N,\tau\mu}|=30\, GeV$,
$|\bar{\xi}^{E}_{N,\tau e}|=0.001\, GeV$
($|\bar{\xi}^{E}_{N,\tau\tau}|=100\, GeV$, $|\bar{\xi}^{E}_{N,\tau
\mu}|=30\, GeV$), $sin\theta_{\tau\mu}=0.5$
($sin\theta_{\tau\tau}=0.5$). Notice that we take $sin\theta_{\tau
e}=0$ ($sin\theta_{\tau\mu}=0$) for the decays $\mu\rightarrow
e\gamma$ ($\tau\rightarrow \mu\gamma$). Here solid (dashed, small
dashed,dotted) line represents the dependence to the coefficient
$|g[p_0,k,k]|$, $|a[p_0]|$, $|b[p_0]|$ and $|e[p_0]|$, in the case
that the other coefficients have the same numerical value
$10^{-20}$. The coefficient $g[p_0,k,k]|$ ($|a[p_0]|$, $|b[p_0]|$
and $|e[p_0]|$) is CP even (odd) and the source of the $A_{CP}$ is
the complex nature of the couplings for the part proportional to
the coefficient $g[p_0,k,k]|$ and the CP odd nature of the
coefficients for the part proportional to  coefficients
$|a[p_0]|$, $|b[p_0]|$ and $|e[p_0]|$ (see eq. \ref{omega2}). For
the $\mu\rightarrow e\gamma$ ($\tau\rightarrow \mu\gamma$) decay
$A_{CP}$ is much more sensitive to the coefficients $g[p_0,k,k]|$
and $|e[p_0]|$ ($g[p_0,k,k]|$, $|e[p_0]|$ and $|a[p_0]|$ )
compared to the ones $|a[p_0]|$ and $|b[p_0]|$ ($|b[p_0]|$). It
increases with the increasing values of $g[p_0,k,k]|$ and
$|e[p_0]|$ ($g[p_0,k,k]|$, $|e[p_0]|$ and $|a[p_0]|$). Notice that
in the case of real couplings the coefficient $g[p_0,k,k]|$ does
not give contribution to the $A_{CP}$  However the numerical value
of $A_{CP}$ is very small, at the order of the magnitude of
$10^{-19}$ for both decays and it seems that it is not possible to
detect even in the future experiments.

At this stage we would like to summarize our results:

We analyse the Lorentz and CPT violating  effects on the $BR$ and
$A_{CP}$ for the LFV decays $\mu\rightarrow e\gamma$ and
$\tau\rightarrow \mu\gamma$ in the framework of the model III.
Here we assume that these effects are only due to the QED part of
the interactions. By taking the numerical values of the
coefficients at the order of the magnitude of $10^{-20}-10^{-18}$
we study the relative behaviors of different coefficients
\begin{itemize}
\item The contribution of the Lorentz and CPT violating part to
the BR of the decays $\mu\rightarrow e\gamma$ ($\tau\rightarrow
\mu\gamma$) is at the order of the magnitude of $10^{-32}$
($10^{-26}$), which is too small to be detected. For the decay
$\mu\rightarrow e\gamma$ the BR is more sensitive to the
coefficient $|e|$ compared to others and for its large values the
BR reaches to the order of $10^{-30}$. For the decay
$\tau\rightarrow \mu\gamma$ the BR is sensitive to the
coefficients $|e|$, $|a|$, $|e|$ and $|b|$ and it can take the
values at the order of the magnitude of $10^{-24}$

\item  We predict the numerical value of $A_{CP}$ at the order of
the magnitude of $10^{-19}$ for both decays. The source of the
$A_{CP}$ is the coefficients $g[p_0,k,k]=m_{l_1}\, g[0,k,k]$,
$a[p_0]=m_{l_1}\,a[0]$, $b[p_0]=m_{l_1}\,b[0]$ and
$e[p_0]=m_{l_1}\,e[0]$. Here $g[0,k,k]$ is CP even and the others,
$a[0],b[0],e[0]$, are CP odd appearing in the lagrangian eq.
(\ref{CPTLag}) (see \cite{Kos5} for details). The $A_{CP}$ is
nonzero due to the complex nature of the couplings for the part
$\Omega_g$ and CP odd nature of the coefficients for the part
$\Omega_a+\Omega_b+\Omega_e$. We observe that the $A_{CP}$ is too
small to be measured even in future experiments.
\end{itemize}
\section{Appendix}
The coefficients $A^{(\prime)}_i(x,y)$, $B^{(\prime)}_{i\,\mu
\alpha}(x,y)$, $C^{(\prime)}_{i\,,\alpha}(x,y)$,
$D^{(\prime)}_{i}(x,y)$, $E^{(\prime)}_{i\,,\mu}(x,y)$ in eqns.
(\ref{Mat2}) and (\ref{Hh}) reads
\begin{eqnarray}
A_i(x,y)\!\!\!\!\!&=&\!\!\!\! \!\Bigg
\{(\frac{1}{L_{2,h^0}}+\frac{1}{L_{2,A^0}})
(1+x+y)\!+\!\!(\frac{1}{L^2_{2,h^0}}+\frac{1}{L^2_{2,A^0}})
f_1+\!\! (\frac{1}{L^2_{2,h^0}}-\frac{1}{L^2_{2,A^0}})\, m_{l_1}
m_i\, x (x+2 y) \Bigg \} a.k \nonumber \\&+& \Bigg \{m_i \,y
\Bigg( (\frac{1}{L^2_{1,h^0}}+\frac{1}{L^2_{1,A^0}})\,f_2+ 2
(\frac{1}{L^2_{1,h^0}}-\frac{1}{L^2_{1,A^0}}) m_{l_1} m_i\, x
\Bigg )+ (\frac{1}{L_{1,h^0}}+\frac{1}{L_{1,A^0}}) m_i\, y
\nonumber \\&-& \Big (( m_i +x\, m_{l_1}) \frac{1}{L^2_{2,h^0}}+(
m_i -x\, m_{l_1})  \frac{1}{L^2_{2,A^0}} \Big) (x+y) (m_i^2-x \,y
\,m^2_{l_1}) - \Bigg( \Big( m_i (-2+x+y)\nonumber \\ &-& 4\, x\,
(x+y)\, m_{l_1} \Big)\,\frac{1}{L_{2,h^0}}+\Big( m_i (-2+x+y)+4
\,x \,(x+y)\, m_{l_1}\Big) \,\frac{1}{L_{2,A^0}}\Bigg) \Bigg \}
e.k \nonumber
\\&+& \!\! (\frac{1}{L^2_{1,h^0}}-\frac{1}{L^2_{1,A^0}})\, m_{l_1}\,
m_i\,y \,(f_3-m_i^2+m^2_{l_1} x (x+y))\, c^{Sym}\nonumber \\&+&
(\frac{1}{L^2_{2,h^0}}-\frac{1}{L^2_{2,A^0}})\, m_{l_1}\, m_i\,x^2
\,(2+x)\, c^{Asym}[p,p^{\prime}] \nonumber \\&+& 2\,i\,x\,
\Bigg(\frac{1}{L^2_{2,h^0}} (m_i \,(x-1)-m_{l_1}\, x\, y)+
\frac{1}{L^2_{2,A^0}} (m_i \,(x-1)+m_{l_1}\, x \,y) \Bigg)\,
h[p,p^{\prime}]\nonumber \\&+& 2\,i\,
(\frac{1}{L^2_{1,h^0}}-\frac{1}{L^2_{1,A^0}}) m_{l_1}\,x^2
\,(x+y)\, (y \,g[p,p^{\prime},k]+x \,g[p,p^{\prime},p])-i\, \Big
(x^2\,(\frac{1}{L_{1,h^0}}-\frac{1}{L_{1,A^0}})\nonumber \\ &+&
x\,(x+1)\,(\frac{1}{L_{2,h^0}}-\frac{1}{L_{2,A^0}}) \Big ) \,
m_{l_1} \,g[k,\beta,\beta]-i \frac{1}{L^2_{2,h^0}} \Bigg(
(x+y)\,x\,\Big( m_i \,(1-x-y)\nonumber \\ &+& 2\, m_{l_1}\, x
\,(1-y)\Big )
 g[p,p^{\prime},k] + \frac{1}{2}\, x^2 \,\Big( m_i\, (2\,
x+y-2)+ 4\, m_{l_1}\, x\, (-1+y)\, \Big)\,
g[p,p^{\prime},p]\,\Bigg)\nonumber
\\&-& i \frac{1}{L^2_{2,A^0}} \Bigg( (x+y)\,x\, \Big(-m_i (x+y-1)+ 2\,
m_{l_1}\, x (-1+y)\Big)\, g[p,p^{\prime},k]\nonumber
\\&+& \frac{1}{2}\,x^2 \,\Big( m_i\, (2\, x+y-2)-4\, m_{l_1}\, x\,
(-1+y)\, \Big)\, g[p,p^{\prime},p]\,\Bigg) \, , \nonumber \\
A^{\prime}_i(x,y)&=& \Bigg \{
(\frac{1}{L^2_{1,h^0}}+\frac{1}{L^2_{1,A^0}})\, f_4 \,(1-x-y)-2\,
(\frac{1}{L^2_{1,h^0}}-\frac{1}{L^2_{1,A^0}})\,x\, (x-1)\, m_i\,
m_{l_1} + \nonumber
\\&+& (\frac{1}{L_{1,h^0}}+\frac{1}{L_{1,A^0}})\, (-1+3 x-y)
-(\frac{1}{L^2_{2,h^0}}-\frac{1}{L^2_{2,A^0}})\, x^2 \,m_i\,
m_{l_1}-(\frac{1}{L^2_{2,h^0}}+\frac{1}{L^2_{2,A^0}})\, f_4
\nonumber \\&+&  (\frac{1}{L_{2,h^0}}+\frac{1}{L_{2,A^0}})
\,(1+x+y) \Bigg \} b.k - i\, \Bigg \{
(\frac{1}{L^2_{1,h^0}}+\frac{1}{L^2_{1,A^0}}) m_i\,y\,
(m_{l_1}^2\, x\, (x+y)-m_i^2) \nonumber \\&-& 3\, m_i\,
(\frac{1}{L_{1,h^0}}+\frac{1}{L_{1,A^0}})- \Big
(\frac{m_i+m_{l_1}\, x}{L^2_{2,h^0}}+\frac{m_i-m_{l_1}\,
x}{L^2_{2,A^0}}\Big) (x+y) (m_i^2+x\, y \,m^2_{l_1})\nonumber
\\&-& \Big(\frac{3\, m_i+4\, x \,m_{l_1}}{L_{2,h^0}}+\frac{3
\,m_i-4\, x\, m_{l_1}}{L_{2,A^0}}\Big)\,(x+y)\, \Bigg \}f.k
\nonumber \\&+& (\frac{1}{L^2_{2,h^0}}-\frac{1}{L^2_{2,A^0}})\,
m_{l_1}\, m_i \,x^2 \,(1-3 x- 3 y)\, d^{Asym}[p,p^{\prime}]
\nonumber \\ &+& \frac{1}{2}\,\epsilon_{\rho\theta\alpha\beta}\,
(\frac{1}{L^2_{1,h^0}}+\frac{1}{L^2_{1,A^0}})\,m_i\, x\,
h[\alpha,\beta]\, k_{\rho}\, p_{\theta}\nonumber
\\&+& \frac{1}{2}\,\epsilon_{\rho\theta\alpha\beta}\,
(\frac{1}{L^2_{2,h^0}}+ \frac{1}{L^2_{2,A^0}})\, m_i\, x
\,(1-x-y)\, (x+y)\, g[\beta,\alpha,k]\,k_{\rho}\,p_{\theta} \, ,
\nonumber \\
B_{i\,\mu \alpha}(x,y)&=& -i \Bigg \{ \frac{1}{4}
\epsilon_{\rho\alpha\beta\mu} \Big (\frac{1}{L^2_{1,h^0}}-
\frac{1}{L^2_{2,A^0}}\Big)\, m_i \,m_{l_1}\, x\,(1-4\, x-4 \,y) +
\frac{1}{L^2_{2,h^0}}\Big(f_4 +2\, m_i\, (m_{l_1}\,x + m_i\,
y))\nonumber \\&+& \frac{1}{L^2_{2,A^0}}(f_4 +2\, m_i\, (-m_{l_1}
\,x + m_i\, y)\Big)+ (\frac{1}{L_{2,h^0}}+\frac{1}{L_{2,A^0}})
\,(-1+3\, x+3\, y) \Bigg \} \, k_{\rho}\, b_{\beta} \nonumber
\\\!\!\!\!&-& \!\!\!\!\Bigg \{ m_i\, m_{l_1}\, y\, \Bigg (\!
(\frac{1}{L^2_{1,h^0}}-\frac{1}{L^2_{1,A^0}}) \,f_3+
(\frac{1}{L^2_{1,h^0}}+\frac{1}{L^2_{1,A^0}})\, m_i\, m_{l_1}
\,x\, \Bigg )+3\, (\frac{1}{L_{1,h^0}}-\frac{1}{L_{2,A^0}})\,
m_i\, m_{l_1} \,y \nonumber \\&+& m_{l_1} \,(x+y)\, (m_i^2+
m^2_{l_1}\, x \,y)\Big( \frac{1}{L^2_{2,h^0}}(m_i+ m_{l_1}\, x)+
\frac{1}{L^2_{2,A^0}}(-m_i+ m_{l_1}\, x) \Big)\nonumber \\&+&
m_{l_1} \Bigg (\frac{1}{L_{2,h^0}} \Big(4\, m_{l_1} \,x
\,(x+y)+m_i\, (1+ 3\, x+3\, y)\Big)+ \frac{1}{L_{2,A^0}} \Big( 4\,
m_{l_1}\, x \,(x+y)\nonumber \\ &-& m_i \,(1+3\, x+3
\,y)\Big)\Bigg ) \Bigg \}\, c^{Asym}[\mu,\alpha] -i \,
\epsilon_{\rho\alpha\beta\mu}\, \Bigg \{2\,
(\frac{1}{L^2_{1,h^0}}-\frac{1}{L^2_{1,A^0}})\, m_{l_1}\, m_i
\,x\, y^2 \nonumber \\ \!\!\!\!&+&\!\!\!\!
(\frac{1}{L^2_{2,h^0}}-\frac{1}{L^2_{2,A^0}})\, m_{l_1}\, m_i\,
x\, (x+y)\,(y-1)\, \Bigg \} \, k_{\rho}\, d^{Asym}[\beta,k] - i
\Bigg \{\, (\frac{1}{L_{2,h^0}}-\frac{1}{L_{2,A^0}}) \,m_{l_1}\,
(x+y)\nonumber \\&-&
(\frac{1}{L^2_{2,h^0}}-\frac{1}{L^2_{2,A^0}})\, m_{l_1} \,f_4-
(\frac{1}{L^2_{2,h^0}}+\frac{1}{L^2_{2,A^0}})\, m_{l_1}\, m_i
\,x^2 \Bigg \}\, h[\mu,\alpha]\nonumber \\\!\!\!\!&+&\!\!\!\!
\frac{1}{2} \epsilon_{\rho\alpha\beta\mu}
(\frac{1}{L^2_{1,h^0}}-\frac{1}{L^2_{1,A^0}}) m_{l_1}\,x \,y^3\,
 k_{\rho}\, g[\beta,k,k] +i \Bigg \{ \frac{1}{L_{1,h^0}}
\Bigg( m_i\,(x-1)-m_{l_1} (x+2\, y+2\, x\, y)\Bigg)\nonumber
\\&+&\frac{1}{L_{1,A^0}}\, \Bigg( m_i\,(x-1)+m_{l_1}\,
 (x+2\, y+2 \,x \,y)\Bigg) \Bigg \}\, g[k,\alpha,\mu] - 2 i
\,(\frac{1}{L_{1,h^0}}-\frac{1}{L_{1,A^0}})\, m_{l_1}\, x\, y
\nonumber \\ &\times& g[p,\alpha,\mu] -\frac{i}{2} \Bigg
(\frac{1}{L_{1,h^0}} \Big( 2 \,m_i \,(x-1)+ m_{l_1}\, (x \,(y-2)-
4 \,y\,(1-3\, y))\Big)\nonumber
\\&+& \frac{1}{L_{1,A^0}} \Big( 2 \,m_i \,(x-1)- m_{l_1}\, (x \,(y-2)-
\!\!\!4\, y\,(1-3\, y))\,\Big) \,\Bigg ) \,g[\mu,\alpha,k]
\nonumber \\ &+& i \frac{1}{2}\,
(\frac{1}{L_{1,h^0}}-\frac{1}{L_{1,A^0}})\, m_{l_1}\, x \,y\,
\Bigg (g[\mu,k,\alpha]-2 \,g[\mu,p,\alpha]+10\, g[\mu,\alpha,p]\,
\Bigg )\, \nonumber \\&-&\!\!\! \frac{i}{4}\,
(\frac{1}{L^2_{2,h^0}}+\frac{1}{L^2_{2,A^0}})\, m_{l_1}^2\, m_i\,
x\, \Bigg ( -4\, x \,(x+y)\, g[k,\alpha,\mu]+ 8\, x\, (x+y)\,
g[p,\alpha,\mu] \nonumber \\&+& 2\, \Big(\,x^2+ y \,(y-1)+ x
\,(2\, y-1)\,\Big) \,g[\mu,\alpha,k]- x\, (-2+2 \,x+y)\,
g[\mu,\alpha,p]\, \Bigg ) \nonumber \\ &+& i
(\frac{1}{L_{2,h^0}}-\frac{1}{L_{2,A^0}})\, m_{l_1}\, x \,(x+y)\,
g[\mu,p,\alpha] +\frac{i}{2}\, \Bigg \{ \frac{1}{L_{2,h^0}} \Bigg(
m_i \,(-1+x+y)+  4 m_{l_1}\,(x\nonumber
\\&+& x^2+y-y^2) \,\Bigg ) +\frac{1}{L_{2,A^0}} \Bigg( m_i \,(-1+x+y)- 4\,
m_{l_1}\,(x+x^2+y-y^2)\, \Bigg ) \Bigg
\} \,g[k,\alpha,\mu]\nonumber \\
&-& \frac{i}{2} \,\frac{1}{L_{2,h^0}} \,\Bigg( m_i \,(-1+x+y)+
m_{l_1}\, (11\, x^2+ 8\, y \,(y-1)+ (19 \,y-8)\,x) \Bigg )
\nonumber \\ &-& \frac{i}{2} \frac{1}{L_{2,A^0}} \Bigg( m_i\,
(-1+x+y)+ m_{l_1}\, (-11 \,x^2- 8 \,y \,(y-1)- (19\, y-8)\, x)
\,\Bigg )\, g[\mu,\alpha,k] \nonumber
\\ &+& \frac{i}{2}
(\frac{1}{L_{2,h^0}}-\frac{1}{L_{2,A^0}})\,\Bigg ( m_{l_1}\, \Big(
4 \,(x+y-1)\, g[p,\alpha,\mu]-(x+y)\, g[\mu,k,\alpha]+ 2 \,(5\,
x+5\, y-4)\nonumber \\
&\times&  g[\mu,\alpha,p] \Big) \Bigg) \, , \nonumber \\
B^{\prime}_{i\,\mu \alpha}(x,y)&=& -i
\epsilon_{\rho\alpha\beta\mu} \, \Bigg \{
-(\frac{1}{L^2_{1,h^0}}-\frac{1}{L^2_{1,A^0}})\, m_{l_1}\, m_i\,
x+(\frac{1}{L^2_{1,h^0}}+\frac{1}{L^2_{1,A^0}}) \,f_3\,
(-1+x+y)\nonumber\\ \!\!\!\!&+& \!\!\!\!
(\frac{1}{L_{1,h^0}}+\frac{1}{L_{1,A^0}}) \, (-1+3\, x+3\, y)+
(\frac{1}{L^2_{2,h^0}}+ \frac{1}{L^2_{2,A^0}}) \Big(\, m_i^2\,
(y-1)+m_{l_1}^2 \,x\, y\, (-1+x+y)\,\Big)\nonumber
\\&+& (\frac{1}{L_{2,h^0}}+\frac{1}{L_{2,A^0}})\,(-1+3 \,x+3 \,y)
\Bigg \} \, k_{\rho}\, a_{\beta} - m_{l_1}^3\, m_i \,x \,\Bigg \{
2 \,(\frac{1}{L^2_{1,h^0}}- \frac{1}{L^2_{1,A^0}}) \,y^2\nonumber
\\&+& (\frac{1}{L^2_{2,h^0}}-\frac{1}{L^2_{2,A^0}})\, (y-1) \,(x+y)\,
\Bigg \} \,d^{ASym}[\mu,\alpha ] \nonumber \\&-& \!\!
\epsilon_{\rho\alpha\beta\mu}\, (\frac{1}{L^2_{1,h^0}}+
\frac{1}{L^2_{1,A^0}})\, m_i \,\Bigg \{  \Big( 2\, (x-1)\, y\,
h[k,\beta]+x \,(1+2 \,x)\, h[p,\beta]\Big) \,k_{\rho}\nonumber
\\&-& x \,\Big( (1+ 2 \,y) \,h[k,\beta]\, p_{\rho}-
h[\beta,\alpha]\, k_{\rho}\, p_{\theta}+ (2\,y +1)\,
h[\mu,\beta]\, k_{\rho} \,p_{\theta} \Big )\,\Bigg \}\nonumber
\\\!\!\!\!&-&\!\!\!\!\!\!  \epsilon_{\rho\theta\alpha\beta}\, \Bigg \{
\frac{1}{L^2_{2,h^0}} \, \Bigg (2\, x \, (m_i+m_{l_1} \,x)\,
h[p,\beta]- (m_i \,x^2+2\, m_i \,y+2\, m_{l_1}\, x^2 \,y+2
\,m_{l_1}\, x \,y^2\,) \,h[k,\beta] \Bigg) \nonumber \\&+&
\frac{1}{L^2_{2,A^0}}\, \Bigg ( 2\, x\, (m_i-m_{l_1}\, x)\,
h[p,\beta]- (m_i\, x^2+2 \,m_i \,y-2 \,m_{l_1} \,x^2 \,y-2
\,m_{l_1}\, x\, y^2) \,h[k,\beta] \Bigg) \Bigg \} \,k_{\rho}
\nonumber \\&-& \epsilon_{\rho\alpha\beta\mu}\, \Bigg \{
(\frac{1}{L^2_{1,h^0}}-\frac{1}{L^2_{1,A^0}})\, (x+y)\, \Big( x
\,g[p,\beta,k]-2\, y \, g[\beta,k,k]-2 \,x\, g[\beta,k,p]-x\,
g[\beta,p,k] \,\Big) \nonumber \\ &+&\frac{1}{2}\, (x+y)\,
m_{l_1}^2\, y^2\, (\frac{1}{L^2_{1,h^0}}+\frac{1}{L^2_{1,A^0}})
\,g[\mu,\alpha,k]\, \Bigg \}\, k_{\rho} +
\frac{1}{2}\,(\frac{1}{L_{1,h^0}}-\frac{1}{L_{1,A^0}})\, m_{l_1}\,
x \,y \nonumber \\ &\times& \Bigg (
\,\epsilon_{\alpha\beta\lambda\mu}\, \Big( g[k,\beta,\lambda]+2\,
g[\lambda,\beta,k]\, \Big) + \epsilon_{\rho\alpha\beta\lambda}\,
\Big(2\, g[\lambda,\beta,\mu]+
g[\mu,\beta,\lambda] \,\Big) k_{\rho} \Bigg ) \nonumber \\
&+& \frac{1}{2} \epsilon_{\rho\alpha\beta\mu}\, (x+y)\, \Bigg(
\frac{1}{L^2_{2,h^0}} \Big( m_i\, (-1+x+y)-4 m_{l_1}(y-1) (x+y)
\Big) \nonumber \\ &+& \frac{1}{L^2_{2,A^0}} \Big(m_i\, (-1+x+y)+4
\, m_{l_1}\, (y-1)\, (x+y)\,  \Big) \Bigg)\, k_{\rho} \,
g[\beta,k,k] \nonumber
\\ &-& \frac{1}{4}\,\epsilon_{\rho\alpha\beta\mu}\,\Bigg \{
\frac{1}{L^2_{2,h^0}}\, \Big( 8 \,m_{l_1}\, x\, (y-1)\, (x+y)\,
(g[p,\beta,k]-g[\beta,k,p])\nonumber \\ &+& m_i\, x\, (-2+2
x+y)\,g[\beta,k,p] \,\Big ) - \frac{1}{L^2_{2,A^0}} \Big(8\,
m_{l_1}\, x\, (y-1)\,(x+y)\, (g[p,\beta,k]-g[\beta,k,p])\nonumber
\\&-& m_i\, x\, (-2+2 x+y)\, g[\beta,k,p]\, \Big ) \Bigg \}\,
k_{\rho} \nonumber
\\&-& \frac{1}{4}\,(\frac{1}{L^2_{2,h^0}}+\frac{1}{L^2_{2,A^0}})\,
m_{l_1}^2\, m_i \,x\, (x+y)\, \Bigg ( 2\, x\,
\epsilon_{\alpha\beta\lambda\mu}\, g[\lambda,\beta,k]\,
+\epsilon_{\rho\alpha\beta\lambda}\, (1+x-y)
\,g[\lambda,\beta,\mu]\, k_{\rho} \Bigg ) \nonumber \\ &-&
\frac{1}{2}\, (\frac{1}{L_{2,h^0}}-\frac{1}{L_{2,A^0}})\, x
\,(x+y)\,
 m_{l_1}\, \Bigg ( (g[k,\beta,\lambda]+2 \, g[\lambda,\beta,k])\,
\epsilon_{\alpha\beta\lambda\mu} +(g[\mu,\beta,\lambda]\nonumber
\\&+& 2\, g[\lambda,\beta,\mu]) \epsilon_{\rho
\alpha\beta\lambda} k_{\rho} \Bigg ) \, , \nonumber \\
C_{i\,,\alpha}(x,y)&=& \Bigg \{ m_i \, y\, \Bigg (
(\frac{1}{L^2_{1,h^0}}+\frac{1}{L^2_{2,A^0}}) \,m_i\, m_{l_1}\, x+
(\frac{1}{L^2_{1,h^0}}-\frac{1}{L^2_{2,A^0}}) \,f_3\, \Bigg) +3\,
m_i \,y\, (\frac{1}{L_{1,h^0}}-\frac{1}{L_{2,A^0}}) \nonumber
\\&+&(x+y)\, (m_i^2+m_{l_1}^2\, x \,y\,)\,\Bigg (
\frac{1}{L^2_{1,h^0}} (m_i+m_{l_1}\,x)+\frac{1}{L^2_{2,A^0}}\,
(-m_i+m_{l_1} \,x) \Bigg )\nonumber \\\!\!\!\!&+& \!\!\!\!
\frac{1}{L_{2,h^0}} \Bigg ( 4\, m_{l_1}\, x \,(x+y)+ m_i \,(1+ 3\,
x+3 \,y)\Bigg)+\frac{1}{L_{2,A^0}} \Bigg ( 4 \,m_{l_1} \,x \,
(x+y)- m_i \,(1+3\,x+3\, y) \Bigg) \Bigg \} \nonumber \\ &\times&
c^{ASym}[k,\alpha] \nonumber \\&-& i \Bigg\{
(\frac{1}{L^2_{2,h^0}}+\frac{1}{L^2_{2,A^0}})\, m_i\, m_{l_1}\,x^2
+ (\frac{1}{L^2_{2,h^0}}-\frac{1}{L^2_{2,A^0}})\, f_4
-(\frac{1}{L_{2,h^0}}-\frac{1}{L_{2,A^0}})\, (x+y) \Bigg\}\,
h[k,\alpha] \nonumber \\ \!\!\!\! &+& \!\!\!\!
\frac{i}{2}\,(\frac{1}{L_{1,h^0}}-\frac{1}{L_{1,A^0}})\, y\, \Big(
2 \,x \,g[p,p^{\prime},\alpha]\!+ \!(5\, x+12\, y)\,
g[k,\alpha,k]\!+\!10\, x \,g[k,\alpha,p]+4 \,x \, g[\alpha,p,k]\,
\Big) \nonumber \\\!\!\!\!&-& \!\!\!\!\frac{i}{4} \,m_i\, m_{l_1}
\,(\frac{1}{L^2_{1,h^0}}+ \frac{1}{L^2_{1,A^0}})\, x \,\Bigg( \,x
\,(-2+2 \,x+y)\, g[k,\alpha,p]-2\, (x+y)\, \Big (
(1+x-y)\,g[\alpha,k,k]\nonumber
\\ &-& 4 x\, g[\alpha,p,k]\, \Big ) \Bigg ) -
i\,(\frac{1}{L_{1,h^0}}-\frac{1}{L_{1,A^0}}) \, x\, \Bigg( (x+y)\,
g[p,p^{\prime}\alpha]+(5\, x+5 \,y-4)\, g[k,\alpha,p] \nonumber \\
&-& \Big( 7 \,x^2+12\, y \,(y-1)+ x (19 \,y-12)\Big)\,
g[k,\alpha,k]- 4 \,x \,(-1+x+y)\, g[\alpha,p,k]
\Bigg ) \Bigg \} \, , \nonumber \\
C^{\prime}_{i\,,\alpha}(x,y)&=& m_i \, m_{l_1}^2 \, x \, \Bigg \{
2 \, (\frac{1}{L^2_{1,h^0}}-\frac{1}{L^2_{1,A^0}}) \,
y^2+(\frac{1}{L^2_{2,h^0}}-\frac{1}{L^2_{2,A^0}})  \,(x+y) \,
(y-1) \,\Bigg \}  \,d^{ASym}[k,\alpha]\nonumber \\ &+& 3 m_i y\,
(\frac{1}{L_{1,h^0}}-\frac{1}{L_{1,A^0}})\,c^{ASym}[k,\alpha]
+\frac{i}{2}\, (\frac{1}{L^2_{1,h^0}}-\frac{1}{L^2_{1,A^0}})\,
m_{l_1}^2 \,x \,y^3\, g[\alpha,k,k]\nonumber \\ &+& \frac{i}{2}\,
(\frac{1}{L_{1,h^0}}-\frac{1}{L_{1,A^0}}) \,\Big( \,2\,x
\,g[p,p^{\prime}\alpha]-(5 \,x+12 \,y)\,g[\alpha,k,k]-10\, x\,
g[\alpha,k,p]-4 \,x\, g[\alpha,p,k]\,\Big) \nonumber \\ &+&
\frac{1}{4}\, \epsilon_{\rho \alpha\theta\beta}\,  m_i \, m_{l_1}
\, (\frac{1}{L^2_{2,h^0}}+\frac{1}{L^2_{2,A^0}})\, \Big( x^2+ y\,
(y-1)+ x \,(2 y-1)\,\Big)\,
g[\beta,\theta,k]\, k_{\rho} \, , \nonumber \\
D_{i}(x,y)&=&\frac{1}{1-x-y} \Bigg(\frac{1}{L_{1,h^0}}\,\Big(
m_i\,(x-1)+m_{l_1}\,x\,(y-1)\Big) - \frac{1}{L_{1,A^0}})\,\Big(
m_i\,(1-x)+m_{l_1}\,x\,(y-1)\Big)\Bigg) \nonumber \\
&-&\Bigg( \frac{1}{L^2_{2,h^0}} \Big(m_i\, (x^2-2\,
x-2 \,y+2\, x\, y)- 2 \,m_{l_1} \,x \,y\, (x+y) \Big) \nonumber \\
&-& \frac{1}{L^2_{2,A^0}} \Big(m_i\, (x^2-2 \,x-2\, y+2 \,x \,y)+
2\, m_{l_1} \,x\, y\,(x+y) \Big)\Bigg)\,
 a.k \nonumber \\
&-& \Bigg( \frac{1}{L^2_{2,h^0}} \Big(2 \,x\, (m_i+m_{l_1}\,
x)\Big)-\frac{1}{L^2_{2,A^0}} \Big(2 \,x \,(m_i-m_{l_1}\, x)\Big)
\Bigg)\, a.p \nonumber\\&+& \frac{1}{4}
(\frac{1}{L^2_{2,h^0}}-\frac{1}{L^2_{2,A^0}})\,m_i\,x (1-4 \,x-4
\,y)\,(b.k-2\, b.p)- \Bigg \{
\frac{f_6}{L^2_{1,h^0}}+\frac{f_5}{L^2_{1,A^0}}\nonumber
\\ &-& 2\, y\, \Big( \frac{1}{L_{1,h^0}}\,(3 \,m_i+4 \,x\,
m_{l_1})-\frac{1}{L_{1,A^0}}\,(3\, m_i-4\, x\, m_{l_1}) \Big)-
\frac{f_8}{L^2_{2,h^0}}+\frac{f_7}{L^2_{2,A^0}}\nonumber
\\&+& 2\,\frac{1}{L_{2,h^0}} (m_i\,(1-3\, x-3 \,y)-m_{l_1}\, x)-2\,
\frac{1}{L_{2,h^0}} (m_i\,(1-3\, x-3 \,y)+m_{l_1}\, x)\Bigg)
c^{Sym} \nonumber \\&+& \Bigg ( 2\,
(\frac{1}{L^2_{1,h^0}}-\frac{1}{L^2_{1,A^0}})\,m_i \,x\, y
+(\frac{1}{L^2_{2,h^0}}-\frac{1}{L^2_{2,A^0}})\,m_i \,x^3 \Bigg)\,
c^{ASym}[p,p^{\prime}] \nonumber \\&-& \Bigg \{ y\,
\frac{1}{L^2_{1,h^0}} \Big( m_i^2\, (-1+x-y)+2 \,m_i\,
m_{l_1}\,x\, (-1+x+y)+m_{l_1}^2\, x \,(x+y)\, (x+y-1)\, \Big )
\nonumber \\ &+& y\, \frac{1}{L^2_{1,A^0}} \Big (m_i^2 \,(1-x+y)+2
\,m_i\, m_{l_1} \,x \,(-1+x+y)-m_{l_1}^2\, x \,(x+y)\, (x+y-1)\,
\Big) \nonumber \\
&-& (\frac{1}{L_{1,h^0}}-\frac{1}{L_{1,A^0}})\,y\, (4\, x+4
\,y-3)+ (x+y)\, \frac{1}{L^2_{2,h^0}} \Big( m_i^2 \,(1-x+y)+2\,
m_i \,m_{l_1}\,x\, y\nonumber \\ \!\!\!\! &+& \!\!\!\!m_{l_1}^2\,
x\, y \, (x+y-1)\Big ) - (x+y) \, \frac{1}{L^2_{2,A^0}}\, (m_i^2
\,(1-x+y)-2\, m_i\, m_{l_1} \, x \,y-m_{l_1}^2 x \,y\, (x+y-1)
)\nonumber\\&+& (x+y) \,(\frac{1}{L_{2,h^0}}-\frac{1}{L_{2,A^0}})
\, (-3+ 4\, x+ 4\, y) ) \Bigg \} e.k - \Bigg\{ x
\frac{1}{L^2_{1,h^0}}\, \Big( m_i^2\, (-1+x-y)\nonumber
\\&-& 2\, m_i\,m_{l_1}\,x + m_{l_1}^2 \,x \,(x+y)\, (-1+x+y)
\Big) +x\, \frac{1}{L^2_{1,A^0}} \,(m_i^2 \,(1-x+y)\nonumber \\
&-& 2\, m_i \,m_{l_1}\, x +m_{l_1}^2\, x \,(x+y)\,(1-x-y)-
(\frac{1}{L_{1,h^0}}- \frac{1}{L_{1,A^0}})\,x \,(4 x+4 y-3)
\nonumber \\&-& x \,\frac{1}{L^2_{2,h^0}} \Big (m_i^2 \,(1-x+y)-2
\,m_i \, m_{l_1}\,x+ m_{l_1}^2\, x\, y \,(x+y-1)\,\Big )+ x\,
\frac{1}{L^2_{2,A^0}}\, \Big( m_i^2 (1-x+y)\nonumber \\ &-& 2\,
m_i \,m_{l_1}\, x +m_{l_1}^2\, x \,y \,(x+y-1) \Big )-x\,
(\frac{1}{L_{2,h^0}}-\frac{1}{L_{2,A^0}}) \,(-3+ 4 \,x+ 4 \,y) )
\Bigg \}\, e.p \nonumber \\&-& 2\,i\, x\,
(\frac{1}{L^2_{2,h^0}}-\frac{1}{L^2_{2,A^0}}) (2\, y \,(y-1)+x\,
(3\, y-1))\, h[p,p^{\prime}]- 2\,i\,
(\frac{1}{L^2_{1,h^0}}-\frac{1}{L^2_{1,A^0}})\, x\, (x+y)\,
\nonumber \\
&\times& (-2\, y^2\, g[p,p^{\prime},k]+ x\, (x-2\, y)\,
g[p,p^{\prime},p])+i\,(\frac{1}{L_{1,h^0}}-\frac{1}{L_{1,A^0}})\,
x^2 \,g[k,\beta,\beta]\nonumber \\ &+& i
(\frac{1}{L^2_{2,h^0}}-\frac{1}{L^2_{2,A^0}}) \Bigg ( \,x \,\Big(
-4\, y \,(x+y)\,(x+y-1)\, g[p,p^{\prime},k]\nonumber
\\ &+& 2\, (-2 y+2 \,y^2+3\, x \,y-x)\, g[p,p^{\prime},p]\Big) \Bigg)
- i\,(\frac{1}{L_{2,h^0}}-\frac{1}{L_{2,A^0}})\, x \,(1-x)\,
g[k,\beta,\beta]\Bigg \} \, , \nonumber \\
D^{\prime}_{i}(x,y)&=& -\Bigg \{ \frac{1}{4} \frac{1}{L^2_{1,h^0}}
\Big( 8 \,m_{l_1} \,x \,y\, (-1+x+y)+ m_i \,(4 \,x^2-4 \,x\, y+8\,
y +x) \Big)\nonumber \\ &+& \frac{1}{4} \frac{1}{L^2_{1,A^0}}
\Big( 8 \,m_{l_1} \,x \,y \,(-1+x+y)- m_i\, (4\, x^2-4 \,x \,y+8\,
y +x) \Big) + \frac{1}{L^2_{2,h^0}} \Big( m_i\, (x^2+2\,
y)\nonumber \\&+& 2\, m_{l_1}\, x \,y \,(x+y) \Big)+
\frac{1}{L^2_{2,A^0}} \Big( -m_i (x^2+2\, y)+2 \,m_{l_1}\, x\, y\,
(x+y)\, \Big) \Bigg \} \, b.k \nonumber \\ &-& \Bigg \{ x\,
\frac{1}{2} \,\frac{1}{L^2_{1,h^0}}\, (-4\, m_{l_1}\, x-m_i\,
(3-4\, y)) + \frac{1}{2} \,x \,\frac{1}{L^2_{1,A^0}}(-4\,
m_{l_1}\, x+ m_i\, (3-4 \,y))\nonumber \\ &-& 2\, x\, \Big (
\frac{1}{ L^2_{2,h^0}} (m_i+m_{l_1}\, x)+ \frac{1}{ L^2_{2,A^0}}
(-m_i+m_{l_1}\, x) \Big) \Bigg \}\, b.p \nonumber
\\&+& \Bigg \{ x \,y\, m_{l_1}^2 \Bigg ( \frac{2}{L^2_{1,h^0}}
( m_{l_1} \,x \,(x+y)- m_i \,(y-x))+\frac{2}{L^2_{1,A^0}} \,(
m_{l_1}\, x \,(x+y)- m_i \,(x-y))\Bigg ) \nonumber \\&-& x\,
(x+y)\, (y-1)\, m_{l_1}^2 \Bigg ( \frac{1}{L^2_{2,h^0}} \,(
m_i-2\, m_{l_1} \,x)-\frac{1}{L^2_{1,A^0}} ( m_i+2 \,m_{l_1}\,
x)\Bigg )\nonumber \\ &-& 8\,
(\frac{1}{L_{1,h^0}}+\frac{1}{L_{1,A^0}})\, m_{l_1}\, x \,y - 2
\,(\frac{1}{L_{2,h^0}}+\frac{1}{L_{2,A^0}})\, m_{l_1} \,x\, (1-4\,
x-4\, y) \Bigg \}\, d^{Sym} \nonumber \\&-& \Bigg \{ 4 \,x y^2
(\frac{1}{L^2_{1,h^0}}-\frac{1}{L^2_{1,A^0}})\, m_i+
(\frac{1}{L^2_{2,h^0}}-\frac{1}{L^2_{2,A^0}})\, m_i \,x \,(3\,
x^2+2 \,(y-1)\, y+x\, (5 \,y-3)) \Bigg \} \nonumber \\&\times&
d^{ASym}[p,p^{\prime}] - i \,x\, y^3\, (\frac{1}{L^2_{1,h^0}}-
\frac{1}{L^2_{1,A^0}}) g[p,p^{\prime},k] \, ,  \nonumber \\
E_{i\, ,\mu}(x,y)&=& m_{l_1}\,\Bigg \{ \frac{1}{L^2_{2,h^0}}
\,\Bigg( -m_i^2\, (1+y)+m^2_{l_1}\, x\, y\, (x+y-1)+m_i\,
m_{l_1}\, x \,(x+2 y)\, \Bigg)\nonumber
\\ &-& \frac{1}{L^2_{1,A^0}}\, \Bigg( m_i^2 \,(1+y)-m^2_{l_1}\, x
\,y \,(x+y-1)-m_i \,m_{l_1}\, x\, (x+2 y) \,\Bigg)  \nonumber
\\&-& (\frac{1}{L_{2,h^0}}+\frac{1}{L_{2,A^0}}) \,
(1+x+y)\,\Bigg \}\, a_{\mu} - i \, \epsilon_{\rho \lambda
\alpha\mu}\, (\frac{1}{L^2_{2,h^0}}-\frac{1}{L^2_{2,A^0}}) m_i
\,x\, p_{\rho}\, p^{\prime}_{\lambda}\, b_{\alpha}\nonumber
\\\!\!\!\!&-&\!\!\!\! \Bigg \{ \frac{1}{L^2_{1,h^0}} \,m_i\, y\,
\Bigg ( m_i^2+m_i\, m_{l_1} \,x-m_{l_1}^2\, x\, (x-y)\, \Bigg
)+\frac{1}{L^2_{1,A^0}} \Bigg (m_i\, m_{l_1}\, x-m_i^2+m_{l_1}^2\,
x \,(x-y) \,\Bigg )\nonumber \\&-& \frac{1}{L_{1,h^0}}\, y \,\Bigg
( m_i+4 \,m_{l_1}\, x-2\, m_{l_1}\, y \,\Bigg
)+\frac{1}{L_{1,A^0}}\, \Bigg (m_i-4 \,m_{l_1}\, x+ 2\, m_{l_1}
\,y\, \Bigg ) \nonumber
\\&-& (x+y) \,\Bigg ( \,\frac{1}{L^2_{2,h^0}} \Big ( 3 \,m_i^3+3
\,m_i^2\, m_{l_1}\, x +3 \,m_{l_1}^3\, x^2\, y+ m_i\, m_{l_1}^2\,
x\,(2-2\, x+y) \Big ) \nonumber \\ &+& \frac{1}{L^2_{2,A^0}} \Big
( -3 \,m_i^3+3 \,m^2_i \,m_{l_1} \,x +3 \,m_{l_1}^3 \,x^2 \,y-
m_i\, m_{l_1}^2\, x(\,2-2 \,x+y) \Big ) \Bigg ) \nonumber
\\&-& \frac{1}{L_{2,h^0}} \Bigg ( m_i \,(1+5 \,x+5\, y)+m_{l_1}
\,(6 \,x^2+4 \,x \,y-2 \,y^2) \Bigg
)+\frac{1}{L_{2,A^0}} \Bigg ( m_i \,(1+5 \,x+5\, y)\nonumber \\
&-& m_{l_1}\, (6 \,x^2+4 \,x \,y+2 \,y^2) \Bigg ) \Bigg \}\,
c^{ASym}[k,\mu] \nonumber \\&-& \Bigg \{2\,m_i\,\Big(
(\frac{1}{L^2_{1,h^0}}-\frac{1}{L^2_{1,A^0}})\,
f_3+(\frac{1}{L^2_{2,h^0}}-\frac{1}{L^2_{2,A^0}}) \,m_i
\,m_{l_1}\, x \Big) \nonumber \\&+& 6\,
(\frac{1}{L_{1,h^0}}-\frac{1}{L_{1,A^0}})\, m_i\, y-
\frac{1}{L^2_{2,h^0}} \Bigg ( 2 \,m^3_i \,(x+y)+ 2\, m_{l_1}\,
m_i^2\, x\, (x+y)+ 2\, m_{l_1}^3 \,x^2 \,y \,(x+y)\nonumber \\ &-&
m_{l_1}^2 \,m_i\, x \,(x^2-2 \,x \,(y-1)- 2\, y^2 \Bigg) -
\frac{1}{L^2_{2,A^0}} \Bigg ( -2 \,m^3_i \,(x+y)+ 2\, m_{l_1}\,
m_i^2 \,x \,(x+y)\nonumber \\ &+& 2\, m_{l_1}^3\, x^2\, y \,(x+y)+
m_{l_1}^2 \,x\, m_i \,(x^2-2\, x \,(y-1)-2 \,y^2 \Bigg) \nonumber
\\&-& 2\, \frac{1}{L_{2,h^0}} \Bigg (\, m_i\, (1+3\, x+3\,
y)+m_{l_1}\, (4\, x^2+4 \,x\, y) \Bigg )+2 \,\frac{1}{L_{2,A^0}}
\Bigg (m_i\, (1+3\, x+3\, y) \nonumber \\ &-& m_{l_1} \,(4 \,x^2+
4 \,x \,y) \Bigg ) \Bigg \}  c^{ASym}[p,\mu]-\Bigg \{  m_{l_1}
\,m_i \,y\, \Bigg ( (\frac{1}{L^2_{1,h^0}}+\frac{1}{L^2_{1,A^0}})
\,f_2\nonumber \\ &+& 2\,
(\frac{1}{L^2_{1,h^0}}-\frac{1}{L^2_{1,A^0}})\, m_{l_1} \,m_i
\,x\, \Bigg ) + (\frac{1}{L_{1,h^0}}+\frac{1}{L_{1,A^0}})\,
m_{l_1}\, m_i \,y - m_{l_1}\Bigg ( \frac{1}{L_{2,h^0}} \Big( m_i
\,(x+y-2)\nonumber \\ &-& 4\, m_{l_1}\,x\, (x+y)\,
\Big)+\frac{1}{L_{2,A^0}} \Big( m_i \, (x+y- 2)+ 4 \,m_{l_1}\, x
\,(x+y)\Big) \Bigg ) \Bigg \}\, e_{\mu} \nonumber \\ &-& i \Bigg
\{ \frac{1}{2} \,\frac{1}{L^2_{1,h^0}}\, (m_i^2\,
(1-x-y)+m_{l_1}^2\, (x^2-x^3+x \,y^2)+m_i\, m_{l_1}\, x )\nonumber
\\ &+& \frac{1}{2}\, \frac{1}{L^2_{1,A^0}}\, (m_i^2
\,(-1+x+y)+m_{l_1}^2\, (-x^2+x^3-x \,y^2)+m_i \,m_{l_1}\, x)
\nonumber \\ &+& \frac{1}{2}
\,(\frac{1}{L_{1,h^0}}-\frac{1}{L_{1,A^0}})\, (x-3 \,y) - 2\,
\frac{1}{L^2_{2,h^0}}\, (m_i^2 \,(y-1)+m^2_{l_1} \,x \,y+ m_i\,
m_{l_1} \,y )\nonumber \\ &-& 2\, \frac{1}{L^2_{2,h^0}}\, (m_i^2
\,(1-y)-m^2_{l_1}\, x\, y+m_i\, m_{l_1}\, y\, )-2 \,
(\frac{1}{L_{2,h^0}}-\frac{1}{L_{2,A^0}})\, (3\, x+2\, y) \Bigg \}
\,h[k,\mu] \nonumber \\&+& i  \Bigg \{ 2\, \frac{1}{L^2_{2,h^0}}
\,(f_4+m_{l_1}\, x\, (m_i+m_{l_1}\, x \,y) )+2\,
\frac{1}{L^2_{2,A^0}} (-f_4+m_{l_1}\, x\, (m_i-m_{l_1}\, x \,y)
)\nonumber\\ &-& 2\, (\frac{1}{L_{2,h^0}}-\frac{1}{L_{2,A^0}})
\,(x+y) \,\Bigg \} h[p,\mu] - \epsilon_{\rho\theta\alpha\mu}\,
(\frac{1}{L^2_{2,h^0}}+\frac{1}{L^2_{2,A^0}}) \,x^2\,
 k_{\rho} \,p_{\theta} \, h[p,\alpha]
\nonumber \\&-& 2\,
i\,(\frac{1}{L^2_{1,h^0}}-\frac{1}{L^2_{1,A^0}})\, m_{l_1}^2\, x\,
(x+y)\, \Bigg( y^2\, g[\mu,k,k]+x \,y \,g[\mu,k,p]+x \,y\,
g[\mu,p,k]\nonumber \\ &+& x^2\, g[\mu,p,p] \Bigg ) +
\frac{i}{2}\,(\frac{1}{L_{1,h^0}}-\frac{1}{L_{1,A^0}}) \Bigg( 4
\,(2+3 \,x)\, y\, g[k,p,\mu] \nonumber \\&+& 4\, y \,(1+x+y)\,
g[\mu,k,k]+4\, x\, (x+y) \,g[\mu,k,p]\nonumber \\ &-& 8 \,(y-3\,
y)\, g[\mu,p,k]+24\, x \,y \,g[\mu,p,p]+2 \,x^2 \,m^2_{l_1}\,
g[\mu,\beta,\beta] \Bigg ) \nonumber
\\&+& \frac{i}{2}\, \frac{1}{L^2_{2,h^0}} \Bigg( (x+y)\,
\Big( m_i\, (-1+x+y)-4 \,m_{l_1} \,(y-1) \,(x+y)\,
g[\mu,k,k]\nonumber
\\ &+& 4\, m_{l_1}\, x \,(y-1) \,( (x+y) \,g[\mu,k,p]-(x+y) \,
g[\mu,p,k]+x\, g[\mu,p,p] ) \Big) \Bigg )\nonumber
\\&+& \frac{i}{2} \,\frac{1}{L^2_{2,A^0}} \Bigg( (x+y) \Big(
m_i\, (-1+x+y)+ 4\, m_{l_1} \,(y-1)\, (x+y)\, g[\mu,k,k]\nonumber
\\ &-& 4\, m_{l_1}\, x \,(y-1)\, ( (x+y)\, g[\mu,k,p]-(x+y)
\,g[\mu,p,k]+x\, g[\mu,p,p] ) \Big) \Bigg ) \nonumber \\ &-& i
(\frac{1}{L_{2,h^0}}-\frac{1}{L_{2,A^0}}) \Bigg( 2 \,(x+3
\,x^2-2\, y^2+2\, y)\, g[p,p^{\prime},\mu]+ 4\, (-2\, x+4
\,x^2-2\, y\nonumber \\ &+& 7 \,x \, y + 3 \,y^2)\, g[\mu,k,k]+2\,
x \,(3-8\, x-7\, y)\, g[\mu,k,p]+ 2\, (4\, x-6 \,x^2+4\, y-9\, x\,
y-4 \,y^2)\nonumber \\ &\times&  g[\mu,p,k]- 4 \,x \,(2-3\, x-3
\,y)\, g[\mu,p,p]- m^2_{l_1}\, x\, (x+1)\, g[\mu,\beta,\beta]
\Bigg)  \, ,
\nonumber \\
E^{\prime}_{i\, ,\mu}(x,y)&=& i \,
\epsilon_{\rho\theta\alpha\mu}\,
(\frac{1}{L^2_{2,h^0}}-\frac{1}{L^2_{2,A^0}})\, m_i \,x \,
k_{\rho}\, p_{\theta}\, a_{\alpha} \nonumber \\&+& \Bigg \{ -\,
m_{l_1} \Bigg ( 2\,(\frac{1}{L^2_{1,h^0}}-\frac{1}{L^2_{2,A^0}})\,
m_{l_1} \,m_i\, (x-1)\, x\, +
(\frac{1}{L^2_{1,h^0}}+\frac{1}{L^2_{2,A^0}})\, f_2\, (-1+x+y)
\Bigg ) \nonumber \\ \!\!\!\!&+& \!\!\!\!
(\frac{1}{L_{1,h^0}}+\frac{1}{L_{1,A^0}}) \,m_{l_1}\, (-1+3 x-y) -
(\frac{1}{L^2_{2,h^0}}+\frac{1}{L^2_{2,A^0}}) \,m_{l_1} \,f_4-
(\frac{1}{L^2_{2,h^0}}-\frac{1}{L^2_{2,A^0}}) \,m^2_{l_1} \,m_i\,
x^2 \nonumber \\&+& (\frac{1}{L_{2,h^0}}+\frac{1}{L_{2,A^0}})\,
m_{l_1} \,(1+x+y) \Bigg \} \,b_{\mu} - \Bigg \{ 2\,
(\frac{1}{L^2_{1,h^0}}-\frac{1}{L^2_{1,A^0}})\, m^2_{l_1}\, m_i
\,x \,y^2 \nonumber \\ &+& 2\,
(\frac{1}{L_{1,h^0}}+\frac{1}{L_{1,A^0}})\, m_{l_1}\, (2 x-y)\, y
- 2\, (\frac{1}{L_{2,h^0}}+\frac{1}{L_{2,A^0}} \,m_{l_1}\, (3\,
x^2+4\, x\, y+y^2) \Bigg \}\, d^{ASym} [k,\mu] \nonumber \\
\!\!\!\! &+& \!\!\!\! \Bigg \{ 4\,
(\frac{1}{L^2_{1,h^0}}-\frac{1}{L^2_{1,A^0}})\, m^2_{l_1}\, m_i\,
x\, y^2+ (\frac{1}{L^2_{2,h^0}}-\frac{1}{L^2_{2,A^0}})\,
m^2_{l_1}\, m_i \,x \,(3 \,x^2+2\, (y-1)\, y \nonumber \\ &+& x\,
(5 y-3)) \Bigg \} \, d^{ASym}[p,\mu]-i \Bigg \{
(\frac{1}{L^2_{1,h^0}}+\frac{1}{L^2_{1,A^0}}) \,m_{l_1}\, m_i\, y
\,(m^2_{l_1}\, x \,(x+y)-m_i^2)\nonumber \\ &-& 3\,
(\frac{1}{L_{1,h^0}}+\frac{1}{L_{1,A^0}})\, m_{l_1}\, m_i \,y -
m_{l_1}\, \Bigg (\frac{1}{L^2_{2,h^0}} \,(m_{l_1}\, x+
m_i)+\frac{1}{L^2_{2,A^0}}\, (-m_{l_1}\, x+ m_i) \Bigg ) \nonumber
\\ \!\!\! &\times& \!\!\! (x+y)\, (m^2_{l_1} \,x \,y+ m^2_i)-
(x+y)\, m_{l_1} \Bigg ( 3\,( \frac{1}{L_{2,h^0}}+
\frac{1}{L_{2,A^0}})\, m_i+ 4\,
(\frac{1}{L_{2,h^0}}-\frac{1}{L_{2,A^0}})\, m_{l_1}\, x \, \Bigg )
\Bigg \}\nonumber \\ &\times& f_{\mu} +  2 \,x
\,\epsilon_{\rho\theta\alpha\mu}\,
(\frac{1}{L^2_{1,h^0}}-\frac{1}{L^2_{1,A^0}}) \,(y
\,h[k,\alpha]+x\, h[p,{\alpha}]) \,k_{\rho}\, p_{\theta} \nonumber
\\&+& \epsilon_{\rho\alpha\beta\mu} \,\Bigg \{
\frac{1}{2} \,(\frac{1}{L^2_{1,h^0}}-\frac{1}{L^2_{1,A^0}})\,
 \Bigg ( m_i^2\, (-1+x-y)+ m_{l_1}^2\, x\,
(x+y)\,(x+y-1)\, \Bigg )\, k_{\rho} \nonumber \\
&-& \frac{1}{2}(\frac{1}{L^2_{1,h^0}}+\frac{1}{L^2_{1,A^0}})\, x
\,m_i\, m_{l_1} \,p_{\rho} - \frac{3}{2}
\,(\frac{1}{L_{1,h^0}}-\frac{1}{L_{1,A^0}})\, (x+y)\, k_{\rho} -
\frac{1}{2}\, \frac{1}{L^2_{2,h^0}}\, \Big ( 2 \,m_i\, m_{l_1}\,x
\nonumber \\ &+& m_i^2\, (y+1)+m_{l_1}^2 \,x\, y \,(-1+x+y) \Big )
\,k_{\rho}  + \frac{1}{2}\, \frac{1}{L^2_{2,A^0}} \,\Big ( -2
\,m_i\, m_{l_1}\, x+m_i^2 \,(y+1)\nonumber \\ &+& m_{l_1}^2\, x\,
y \,(-1+x+y) \,\Big ) \, k_{\rho}-\frac{3}{2}\,
(\frac{1}{L_{2,h^0}}-\frac{1}{L_{2,A^0}}) \,(x+y) k_{\rho}\, \Bigg
\}\, h[\alpha,\beta] \nonumber \\ &-& i
(\frac{1}{L^2_{1,h^0}}-\frac{1}{L^2_{1,A^0}})\, m_{l_1}^2\, x
\,y^3 \,g[\mu,p,k]+ \epsilon_{\rho\alpha\beta\mu} \Bigg(
\frac{1}{2}\,(\frac{1}{L_{1,h^0}}-\frac{1}{L_{1,A^0}}) \,x\, y
\,(g[k,\alpha,\beta]\nonumber \\ \!\!\!\! &+& \!\!\!\! 2\,
g[\beta,\alpha,k]) - \frac{1}{2}
\,(\frac{1}{L^2_{2,h^0}}+\frac{1}{L^2_{2,A^0}}) \,m_i\, m_{l_1}\,
x^2 \,(x+y)\, g[\beta,\alpha,k] \nonumber \\ &-&
\frac{1}{2}(\frac{1}{L_{2,h^0}}-\frac{1}{L_{2,A^0}})\, x\,
(x+y)\,(g[k,\alpha,\beta]+2\, g[\beta,\alpha,k]) \Bigg )\,
k_{\rho} \, , \nonumber \\  \label{func1}
\end{eqnarray}
where
\begin{eqnarray}
L_{1,S}&=&m_S^2\,(x+(1-x)\,z_h+x\,(-1+x+y)\,z_{1,h})\nonumber
\, ,\\
L_{2,S}&=&m_S^2\,(x+(1-x)\,z_h-x\, y\, z_{1,h})\nonumber \, ,\\
f_1&=&m_i^2\,(y+1)-m^2_{l_1}\,x\,y\, (x+y-1)\nonumber \, ,\\
f_2&=&m_i^2+m^2_{l_1}\,x\,(x+y)  \nonumber \, , \\
f_3&=&m_i^2-m^2_{l_1}\,x\,(x+y)  \nonumber  \, , \\
f_4&=&m_i^2\,(1-y)+m^2_{l_1}\,x\,y\, (x+y-1)\nonumber
\, , \\
f_5&=&m_i\,(f_3+m_i^2)\,y+m_{l_1}\,x\,
\Big(2\,m_{l_1}^2\,x\,(y-1)\,(x+y)-2\,m_i^2\,y-m_i\,m^2_{l_1}\,
(2\,x^2+y^2+x\,y-y) \Big)\nonumber
\, , \\
f_6&=&-m_i\,(f_3+m_i^2)\,y+m_{l_1}\,x\,
\Big(2\,m_{l_1}^2\,x\,(y-1)\,(x+y)-2\,m_i^2\,y+m_i\,m^2_{l_1}\,
(2\,x^2+y^2+x\,y-y) \Big)\nonumber
\, , \\
f_7&=&2\,m^3_{i}\,(x+y)-2\,m^3_{l_1}\,x^3+m_{l_1}\,m_i\,x\,(-2\,m_i\,
(x+y)+m_{l_1}\,(x+2\,x\,y+2\,y^2-2\,y))\nonumber
\, , \\
f_8&=&2\,m^3_{i}\,(x+y)+2\,m^3_{l_1}\,x^3+m_{l_1}\,m_i\,x\,(2\,m_i\,
(x+y)+m_{l_1}\,(x+2\,x\,y+2\,y^2-2\,y)) \, , \label{funct1}
\end{eqnarray}
with  $z_h=\frac{m_i^2}{m^2_S}$, $z_{1,h}=\frac{m_{l_1}^2}{m^2_S}$
and for $S=h^0,A^0$. Here we use $h[r,r']=h_{\alpha\beta}\,
r^{\alpha}\, r'^{\beta}$, $g[r,r',r'']=g_{\alpha\beta\gamma}\,
r^{\alpha}\, r'^{\beta}\, r''^{\gamma}$ and parametrize the
coefficients $c[r,r']$ and $d[r,r']$ as
$c[r,r']=c^{Asym}_{\alpha\beta}\, r^{\alpha}\,
r'^{\beta}+c^{Sym}\, r.r'$, $d[r,r']=d^{Asym}_{\alpha\beta}\,
r^{\alpha}\, r'^{\beta}+d^{Sym}\, r.r'$, where $Asym$ ($Sym$)
denotes the asymmetry (symmetry) in the indices. Here
$h_{\alpha\beta}\, (g_{\alpha\beta\gamma})$ is taken antisymmetric
with respect to indices (first two indices) and $d_{\alpha\beta}$
$(c_{\alpha\beta})$ traceless.
\section{Acknowledgement}
This work has been supported by the Turkish Academy of Sciences in
the framework of the Young Scientist Award Program.
(EOI-TUBA-GEBIP/2001-1-8)
\newpage
\
\end{document}